\documentclass[
    aps,
    prl,
    twocolumn,
    10pt,
    a4paper,
    superscriptaddress,
    notitlepage,
    longbibliography,
    floatfix
]{revtex4-2}

\usepackage{amsmath}
\usepackage{amssymb}
\usepackage{amsthm}
\usepackage{blindtext}
\usepackage{csquotes}
\usepackage{microtype}
\usepackage{nicefrac}
\usepackage{orcidlink}
\usepackage{pifont}
\usepackage{physics}
\usepackage{siunitx}
\usepackage{textcomp}
\usepackage{txfonts}
\usepackage[capitalize]{cleveref}

\usepackage{graphicx}
\usepackage{float}
\usepackage{xcolor}
\usepackage{color, colortbl}
\usepackage{placeins}

\usepackage{capt-of} 
\usepackage{booktabs}
\usepackage{tabularx}
\usepackage{multirow}

\usepackage{hyperref}
\hypersetup{colorlinks=true, citecolor=blue, urlcolor=blue, linkcolor=blue}

\newcommand{\vect}[1]{\boldsymbol{#1}} 
\newcommand{\uvect}[1]{\hat{\vect{#1}}} 

\newcommand{\e}{\mathrm{e}} 
\newcommand{\iu}{\mathrm{i}} 

\begin{document}
\title{D-Wave Phonon Angular Momentum Texture in Altermagnets by Magnon-Phonon Hybridization}

\author{Hannah Bendin\orcidlink{0009-0006-2155-2834}}
\affiliation{Institute of Physics, Martin Luther University Halle-Wittenberg and Halle-Berlin-Regensburg Cluster of Excellence \enquote{Center for Chiral Electronics}, D-06120 Halle (Saale), Germany}
\affiliation{Institute of Physics, Otto von Guericke University Magdeburg, D-39106 Magdeburg, Germany}
\author{Alexander Mook\orcidlink{0000-0002-8599-9209}}
\affiliation{Institute of Solid State Theory, University of Münster, D-48149 Münster, Germany}
\affiliation{Institute of Physics, Johannes Gutenberg University Mainz, D-55128 Mainz, Germany}
\author{Ingrid Mertig\orcidlink{0000-0001-8488-0997}}
\affiliation{Institute of Physics, Martin Luther University Halle-Wittenberg and Halle-Berlin-Regensburg Cluster of Excellence \enquote{Center for Chiral Electronics}, D-06120 Halle (Saale), Germany}
\author{Robin R.~Neumann\orcidlink{0000-0002-9711-3479}}
\email[Correspondence email address: ]{rneumann@uni-mainz.de}
\affiliation{Institute of Solid State Theory, University of Münster, D-48149 Münster, Germany}
\affiliation{Institute of Physics, Johannes Gutenberg University Mainz, D-55128 Mainz, Germany}
\affiliation{Institute of Physics, Martin Luther University Halle-Wittenberg and Halle-Berlin-Regensburg Cluster of Excellence \enquote{Center for Chiral Electronics}, D-06120 Halle (Saale), Germany}

\begin{abstract}
In altermagnets, the magnon bands are anisotropically spin split in reciprocal space without relativistic or dipolar spin-spin interactions.
In this work, we theoretically study magnons and phonons coupled by spin-lattice interaction in a two-dimensional square-lattice $d$-wave altermagnet.
We show that phonon-chirality-selective magnon-phonon hybridization can be caused by interfacial Dzyaloshinskii-Moriya interaction leading to the emergence of hybrid quasiparticles that possess finite phonon angular momentum.
These hybrid quasiparticles are called magnon polarons and consist of spin-polarized magnons and chiral phonons.
Their phonon angular momentum texture follows the $d$-wave character of the magnon spin texture opening up the possibility of phononic counterparts to the electronic response effects in altermagnets, such as a \emph{phonon angular momentum splitter effect}, i.e., the generation of a transverse phonon angular momentum current induced by a temperature gradient -- the bosonic analog of the spin-splitter effect.
\end{abstract}

\date{\today}
\maketitle

\paragraph{Introduction.}
Chiral phonons~%
\cite{%
    zhangChiralPhononsHighSymmetry2015,%
    zhuObservationChiralPhonons2018,%
    wang_chiral_2024%
},
the quasiparticles of circularly polarized lattice vibrations, have recently been investigated due to a range of emerging phenomena.
They have been found to contribute to spin relaxation \cite{garaninAngularMomentumSpinphonon2015,nakaneAngularMomentumPhonons2018}, ultrafast demagnetization \cite{tauchertPolarizedPhononsCarry2022,mrudul_generation_2025} and the Einstein-de Haas effect \cite{zhangAngularMomentumPhonons2014,dornesUltrafastEinsteinHaas2019,zhang_measurement_2025}.
Notably, chiral phonons carry nonzero angular momentum and can therefore function as angular momentum carriers in nonmagnetic materials \cite{parkPhononAngularMomentum2020,anCoherentLongrangeTransfer2020}.
Here we are using the term \enquote{chiral phonon} synonymously to \enquote{axial phonon} proposed in Ref.~\cite{juraschek_chiral_2025} for phonons that carry angular momentum.

The systems in which chiral phonons occur still require extensive research.
Chiral phonons may, for example, be found in lattices with broken inversion symmetry \cite{zhangChiralPhononsHighSymmetry2015, chenChiralPhononsKagome2019, cohClassificationMaterialsPhonon2023,zhuObservationChiralPhonons2018,ishitoTrulyChiralPhonons2023}.
Due to the preserved time-reversal symmetry, the phonon angular momentum has an odd parity:
$
    \vect{L}(\vect{k})
    =
    -\vect{L}(-\vect{k})
$~\cite{cohClassificationMaterialsPhonon2023}.
In order to realize \emph{even-parity} textures, time-reversal symmetry must be broken~\cite{cohClassificationMaterialsPhonon2023}, e.g., by the coupling to magnons, the quasiparticles of spin excitations~%
\cite{
    ma_antiferromagnetic_2022,%
    cuiChiralitySelectiveMagnonphonon2023,%
    wangRegulationMagnonphononCoupling2023,%
    mella_chiral_2024,%
    ning_spontaneous_2024,%
    wang_magnetic_2024,%
    weissenhoferAtomisticSpinDynamics2024%
}.

\begin{figure}[h!]
    \centering
    \includegraphics[width=\linewidth]{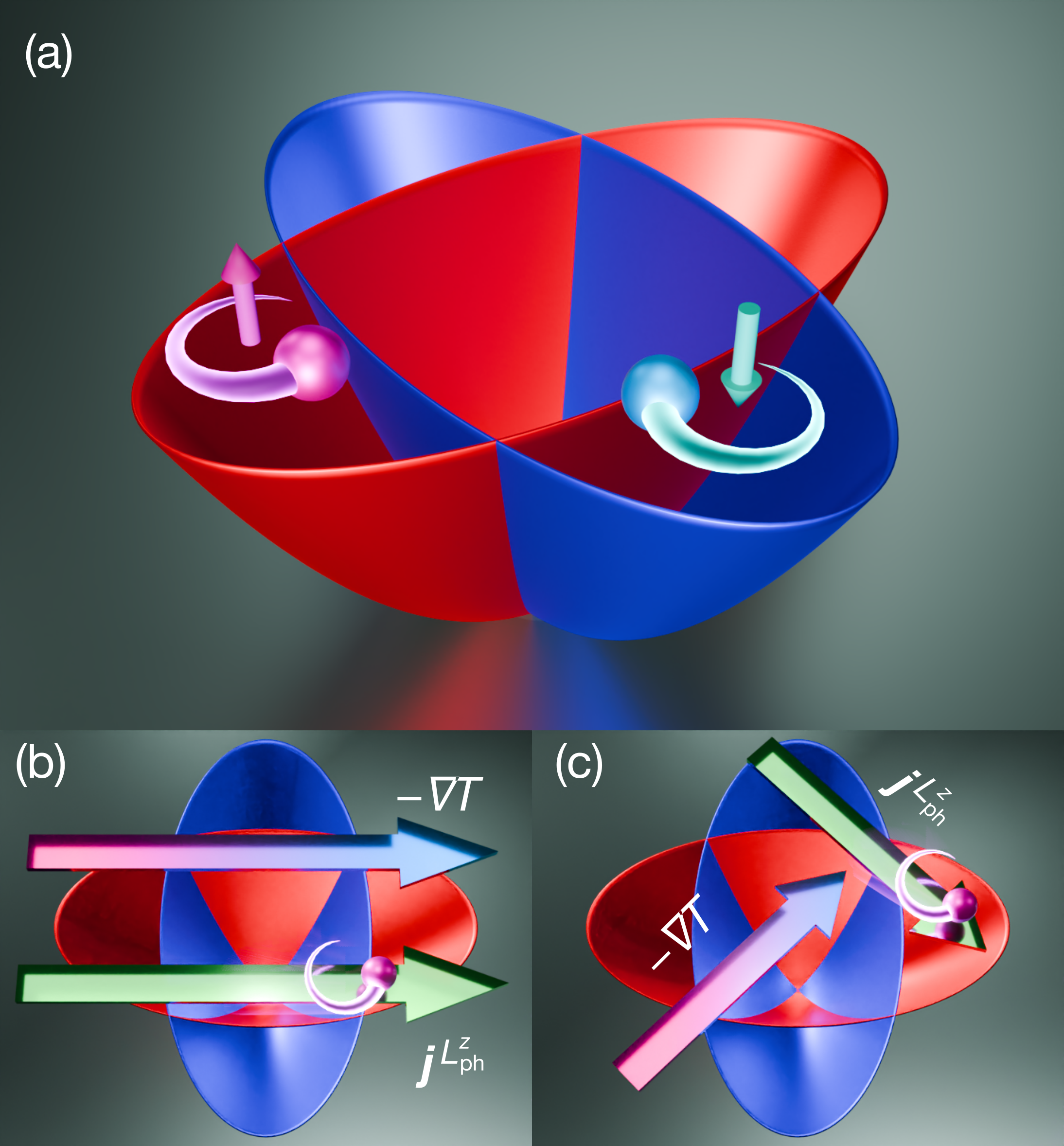}
    \caption{%
        (a) Generation of phonon angular momentum in altermagnets through phonon coupling with the underlying spin-split magnon band structure.
        Their hybridization selectively depends on the circular polarization of the phonons and, thus, it separates left- and right-handed phonons imprinting an alternating phonon angular momentum texture in reciprocal space.
        (b), (c) Visualization of the phonon angular momentum splitter effect, i.e., the generation of a phonon angular momentum current by the application of a temperature gradient.
    }
    \label{fig:graphical_abstract}
\end{figure}

In this work, we delineate an approach to design even-parity phonon angular momentum textures beyond $s$ wave.
Our approach is based on a chirality-selective magnon-phonon coupling, which leads to the hybridization between magnons and phonons.
In particular, we show that the strength of magnon-phonon coupling originating from interfacial Dzyaloshinskii-Moriya interaction (DMI) depends on the magnon spin and the phonon angular momentum.
We study their hybridization in the context of a $d$-wave altermagnet~\cite{smejkalConventionalFerromagnetismAntiferromagnetism2022,smejkalEmergingResearchLandscape2022} and demonstrate how the symmetry of the magnon spin texture can be imprinted onto the phonon angular momentum texture of the emergent magnon polarons [cf.~\cref{fig:graphical_abstract}(a)].
Our findings encourage the exploration of the vibrational degrees of freedom in novel unconventional magnetic phases such as altermagnets or antialtermagnets~\cite{jungwirth_altermagnetism_2025} and highlight the potential of unconventional magnetism for realizing chiral phonons.

\paragraph{Results.}

\begin{figure}
    \centering
    \includegraphics[width=\linewidth]{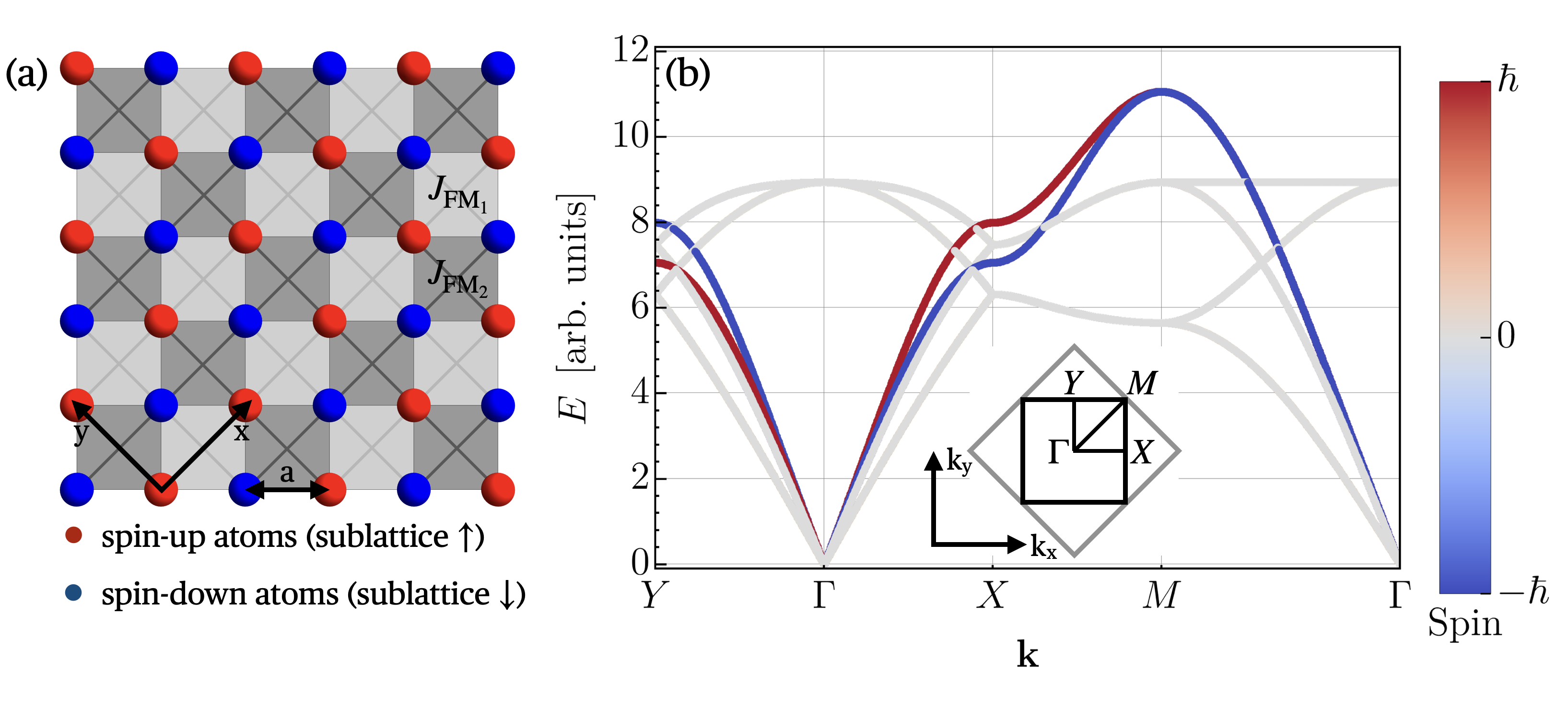}
    \caption{%
        (a) Altermagnetic model comprising antiferromagnetic nearest-neighbor interactions and directional, ferromagnetic next-nearest neighbor interactions as indicated by the gray checkerboard pattern.
        (b) Magnon and phonon modes in the absence of spin-lattice coupling.
        The color indicates the magnon spin.
        Phonon modes are depicted in gray while the red and blue colors confirm the spin-dependent splitting of the magnon modes.
        Inset: Brillouin zone of the magnetic (structural) unit cell shown in black (gray).
        (The coordinate systems of the real and reciprocal spaces are rotated by \SI{45}{\degree} to each other.)
        The parameters used for the calculations are
        $J_{\mathrm{AFM}} = -1$,
        $J_{\mathrm{FM}_1} = 0.383$,
        $J_{\mathrm{FM}_2} = 0.5$,
        $S = 1$,
        $M = 10$,
        $K_{\mathrm{L}}^{(1)} = 160$,
        $K_{\mathrm{T}} = K_{\mathrm{L}}^{(2)} = 40$,
        and
        $a = 1$.
    }
    \label{fig:dispersion_relation_and_lattice}
\end{figure}

We consider the minimal, two-dimensional altermagnet depicted in \cref{fig:dispersion_relation_and_lattice}(a) with the two sublattices $\uparrow$ (red circles) and $\downarrow$ (blue circles).
The localized spins assume a collinear Néel order and interact via antiferromagnetic nearest and anisotropic ferromagnetic next-nearest neighbor Heisenberg coupling (cf.~End Matter).
Using linear spin-wave theory~\cite{supplement}, we obtain the dispersion relation depicted as red and blue lines in \cref{fig:dispersion_relation_and_lattice}(b), which features spin-split magnon modes.
Their splitting is maximal and opposite along the high-symmetry lines $\overline{\Gamma \mathrm{X}}$ and $\overline{\Gamma \mathrm{Y}}$ and vanishes along the nodal lines $\overline{\Gamma \mathrm{M}}$.
Thus, the model exhibits the characteristic features in the magnon band structure expected in altermagnets \cite{smejkalChiralMagnonsAltermagnetic2023,weissenhoferAtomisticSpinDynamics2024,jostChiralAltermagnonMnTe2025}.

The phonon Hamiltonian of the system respects the symmetry of the square lattice and comprises nearest-neighbor interactions with restoring forces along longitudinal $K_{\mathrm{L}}^{(1)}$ and transverse $K_{\mathrm{T}}$ directions with respect to the bonds as well as longitudinal next-nearest neighbor interactions $K_{\mathrm{L}}^{(2)}$ (cf.~End Matter).
The phonon dispersion relation is depicted in gray in \cref{fig:dispersion_relation_and_lattice}(b) along with the magnon modes.
Note that in the structural unit cell the phonon system comprises only a single atom per unit cell forming two acoustic bands.
However, since the magnetic unit cell is larger, the corresponding Brillouin zone is smaller and the bands have to be down-folded yielding four phonon branches.
In general, the checkerboard geometry invoked for the magnons also modifies the phonons, but it does not qualitatively change the results presented here~\cite{supplement}.
Because of the restriction to in-plane displacements, the phonon angular momentum is aligned with the $z$ axis and can be expressed as~\cite{zhangAngularMomentumPhonons2014}
\begin{align}
    L_{\mathrm{ph}}^z
    =
    \sum_{i} \left( u_{i}^x p_{i}^y - u_{i}^y p_{i}^x \right)
    \ .
    \label{eq:angular momentum operator}
\end{align}
Symmetry demands that the expectation values of $L_{\mathrm{ph}}^z$ vanish in an inversion- and time-reversal-symmetric (nonmagnetic) square lattice \cite{cohClassificationMaterialsPhonon2023}.

Turning to the magnon-phonon coupling, the nonrelativistic altermagnet does not exhibit magnon-phonon hybridization because exchange magnetostriction preserves the magnon spin.
To break spin conservation, we additionally invoke DMI, which is the leading-order effect of spin-orbit interaction \cite{dzyaloshinskyThermodynamicTheoryWeak1958,moriyaAnisotropicSuperexchangeInteraction1960}.
In particular, we study interfacial DMI
$
    \hat{H}_{\mathrm{DMI}}
    \propto
    \vect{D}_{ij} \cdot \left( \vect{S}_i \times \vect{S}_j \right)
$,
which results from the breaking of the out-of-plane mirror symmetry.
Since the orientation of the DM vector $\vect{D}_{ij} = D \uvect{R}_{ij} \times \uvect{z}$ is in the plane and perpendicular to the bond vector $\vect{R}_{ij}$, it is also perpendicular to the Néel vector and does not modify the noninteracting magnon dispersion relation~\cite{zhangThermalHallEffect2019}.
(%
    Although the true ground state in the presence of DMI would be a noncollinear state~\cite{dos_santos_modeling_2020}, an easy-axis anisotropy would stabilize the collinear ground state for sufficiently small $D$.
    We do not implement the anisotropy for simplicity, since it does not change the qualitative effects discussed in the following~\cite{supplement}.%
)
The influence of the lattice excitations on the magnons enters the model by the modification of the bond vectors $\vect{R}_{ij} \mapsto \vect{R}_{ij} + \vect{u}_{ij}$.
Expansion of the DMI up to first order in the displacements of the atoms yields the coupling Hamiltonian~%
\cite{
    zhangThermalHallEffect2019,%
    ma_antiferromagnetic_2022,%
    wangRegulationMagnonphononCoupling2023,%
    supplement%
}
\begin{align}
    \hat{H}_{\mathrm{mpc}}
    =
    \frac{D S}{R} \sum_{\langle\langle i,j \rangle\rangle}
    \left\{
    \pm ( \vect{u}_i - \vect{u}_j ) \cdot ( \vect{S}_i - \vect{S}_j )
    \right\}
    \ .
    \label{eq:mpc}
\end{align}
The sum in the expression of $\hat{H}_{\mathrm{mpc}}$ runs over all next-nearest neighbors $\langle \langle i, j \rangle \rangle$.
The \enquote{$+$} sign holds for those sites $i, j$ of the $\uparrow$ and the \enquote{$-$} sign for the $\downarrow$ sublattice.
We note in passing that we have neglected contributions that originate from the distance dependence of $D(R)$~\cite{zhangThermalHallEffect2019}, which we explicitly analyze in the Supplemental Material~\cite{supplement}.

\begin{figure}
    \centering
    \includegraphics[width=\linewidth]{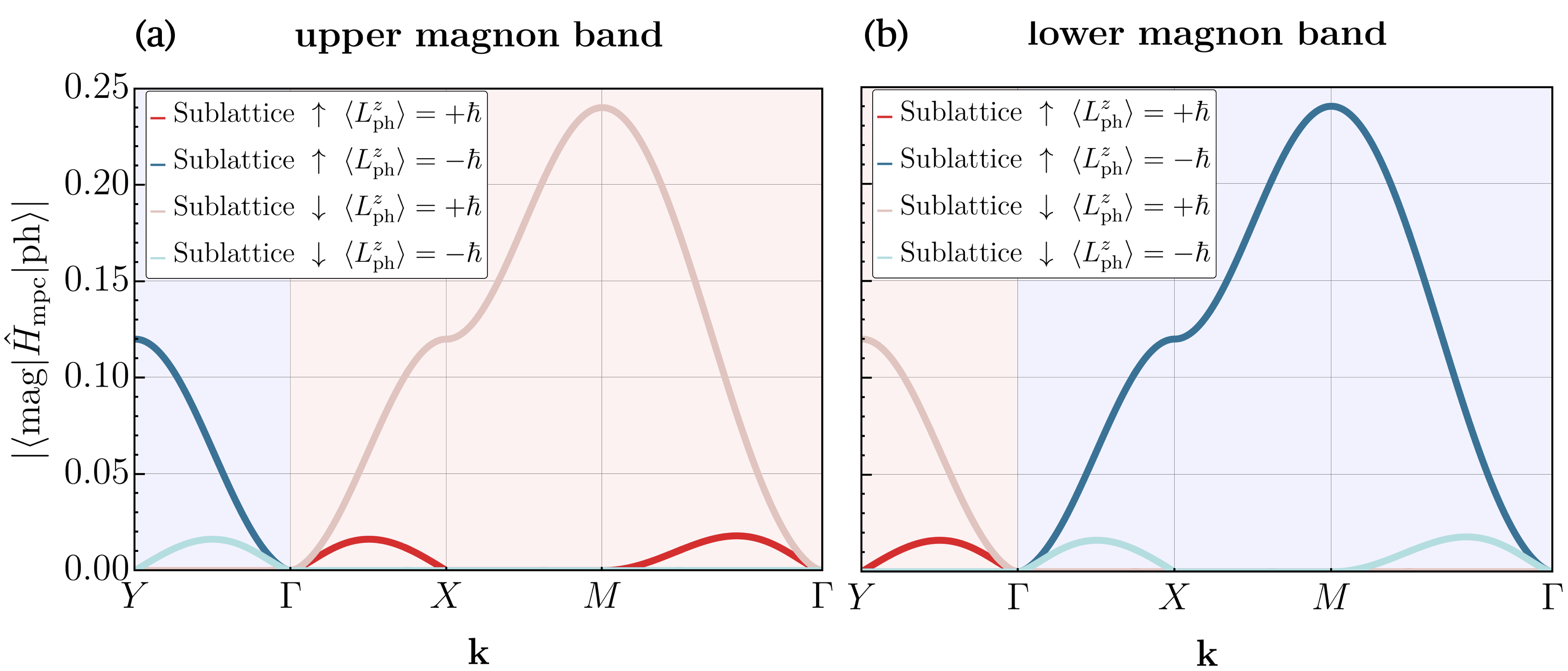}
    \caption{%
        Magnon-phonon coupling strength between the (a) upper, (b) lower magnon branch and the circularly and sublattice-polarized phonon modes.
        All lines represent the absolute value of one of the 8 matrix elements of $\hat{H}_{\mathrm{mpc}}$ [\cref{eq:mpc}] represented in the basis of the magnon bands and the sublattice-resolved chiral phonons (cf.~Supplemental Material~\cite{supplement}).
        The background color and the line color indicate the signs of the magnon spin and the phonon angular momentum, respectively.
        The matching of colors indicates the chiral selectivity of the magnon-phonon coupling $\hat{H}_{\text{mpc}}$.
    }
    \label{fig:coupling_strength}
\end{figure}

To characterize $\hat{H}_{\mathrm{mpc}}$, we project it onto the magnon bands as eigenmodes of the bare magnon Hamiltonian (lower magnon band, upper magnon band) and sublattice-localized chiral phonons (right- and left-handed chiral phonons localized on sublattices $\uparrow$ and $\downarrow$) and take the absolute value of the matrix elements representing all eight interaction channels.
The mathematical details of this analysis are described in the Supplemental Material~\cite{supplement} and the result is plotted in \cref{fig:coupling_strength}.
The background and the line colors indicate the sign of the magnon spin and phonon angular momentum, respectively.
For the upper magnon band [\cref{fig:coupling_strength}(a)], the magnon spin is negative along $\overline{\Gamma\mathrm{Y}}$ (indicated by blue background color), and the band couples exclusively to left-handed circularly polarized phonons with negative angular momentum ($\langle L_{\mathrm{ph}}^z \rangle = -\hbar$; indicated by blue line color) on both sublattices.
The coupling strength on the $\downarrow$ sublattice (faint color) is weaker than on the $\uparrow$ sublattice (vivid color) and vanishes at the Y point.
Along $\overline{\Gamma \mathrm{X}}$ the magnon spin reverses sign (red background color), such that the band couples solely to right-handed circularly polarized phonons with positive angular momentum ($\langle L_{\mathrm{ph}}^z \rangle = \hbar$; red line color).
Along $\overline{\Gamma \mathrm{M}}$, the two magnon branches become degenerate, rendering their labeling arbitrary.
Overall, $\hat{H}_{\mathrm{mpc}}$ couples each magnon spin channel selectively to a single phonon circular polarization, determined by the sign of the magnon spin.

The physical origin of this selectivity can be intuitively understood from \cref{eq:mpc}.
In antiferromagnets and altermagnets the two magnon branches have opposite chiralities, corresponding to clockwise or counter-clockwise spin precession in the $xy$ plane~\cite{rezende_introduction_2019,smejkalChiralMagnonsAltermagnetic2023}.
The dynamic spin precession induces an effective field acting on the lattice displacements through $\hat{H}_{\mathrm{mpc}}$.
For corotating chiral phonons---those exhibiting the same sense of rotation as the spin precession---the time average of \cref{eq:mpc} remains finite leading to an energy shift.
In contrast, the coupling for counter-rotating phonons---those exhibiting the opposite sense of rotation as the spin precession---averages to zero explaining intuitively the chiral selection rule of the coupling.
Linearly or elliptically polarized phonons should be regarded as superpositions of left- and right-handed circular polarizations, which therefore couple to both magnon bands and the coupling strength depends on the proportion of the compatible circular polarization.
Only for chiral phonon modes featuring the incompatible circular polarization, the coupling is zero.
Within a given magnon band, the sublattice amplitudes of the spin precession are generally unequal.
For instance, the spin-down magnon mode exhibits a larger precession amplitude on the $\uparrow$ sublattice, as the magnon spin is defined as the difference of the spin expectation values between the excited and the ground states~\cite{okuma_magnon_2017}. 
As $\vect{k}$ approaches the boundary of the first Brillouin zone, the amplitude of the $\downarrow$ sublattice diminishes, while that of the $\uparrow$ sublattice remains finite.

\begin{figure*}
    \centering
    \includegraphics[width=\linewidth]{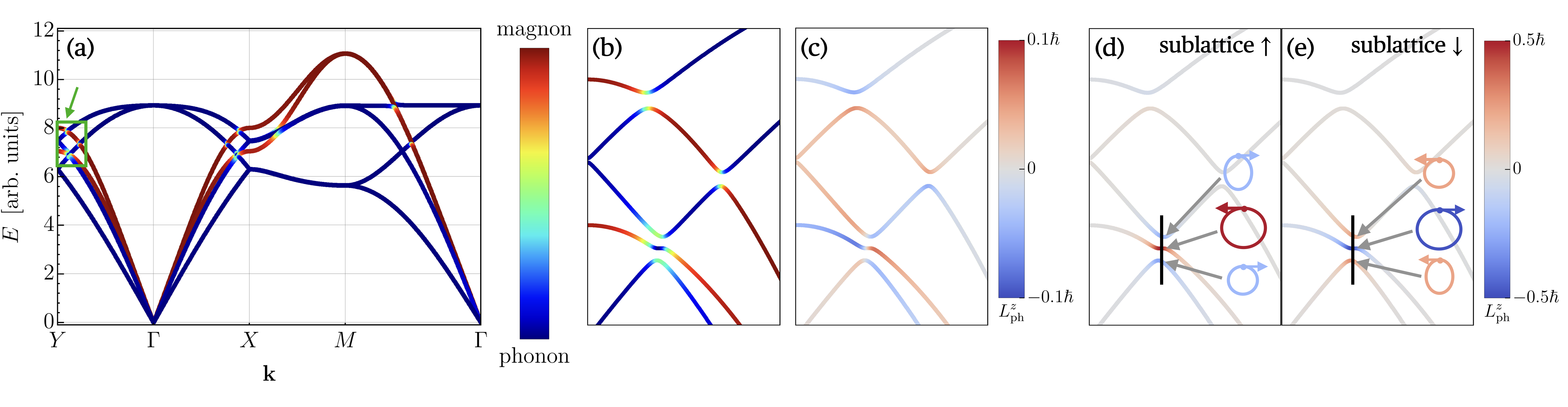}
    \caption{%
        Coupled magnon-phonon hybrid system.
        (a) Band structure of the magnon polarons along a high-symmetry path in the magnetic Brillouin zone.
        The color indicates the magnon/phonon character of the quasiparticles.
        The region highlighted by the green box and arrow is displayed in greater detail in (b)-(e).
        (b) Zoom of the avoided crossing pattern and quasiparticle character indicated by color.
        (c) Illustration of the same region but depicting the phonon angular momentum.
        (d),~(e) Sublattice-resolved phonon angular momentum for sublattice $\uparrow$ and sublattice $\downarrow$, respectively.
        The insets demonstrate the real-space vibrational motion (amplitude and phase) of the respective sublattices.
        The parameters used for the calculations are
        $J_{\mathrm{AFM}} = -1$,
        $J_{\mathrm{FM}_1} = 0.383$,
        $J_{\mathrm{FM}_2} = 0.5$,
        $S = 1$,
        $M = 10$,
        $K_{\mathrm{L}}^{(1)} = 160$,
        $K_{\mathrm{T}} = K_{\mathrm{L}}^{(2)} = 40$,
        $D = 0.25$,
        and
        $a = 1$.
    }
    \label{fig:hybridized_system}
\end{figure*}
The dispersion relation of the coupled system
\begin{align}
    \hat{H} = \hat{H}_{\mathrm{m}} + \hat{H}_{\mathrm{ph}} + \hat{H}_{\mathrm{mpc}}
\end{align}
is plotted in \cref{fig:hybridized_system}(a).
Here, the color indicates the magnon/phonon/mixed character of the quasiparticles.
Evidently, avoided crossings emerge at the intersection of magnon and phonon bands, giving rise to hybrid magnetic and structural excitations dubbed \enquote{magnon polarons}~\cite{li_advances_2021}.
Furthermore, the close-up shown in \cref{fig:hybridized_system}(b) into the highlighted section as part of the $\overline{\Gamma \mathrm{Y}}$ line provides insights into the generation of phonon angular momentum by magnon-phonon coupling through time-reversal symmetry breaking.
In agreement with previous studies, a distinct avoided crossing pattern emerges at the intersection of two phonon bands and a single magnon band~\cite{weissenhoferTrulyChiralPhonons2024,wangRegulationMagnonphononCoupling2023}.
While the energy and phonon character of the central mode remain unchanged, two magnon-polaron modes emerge whose energies are shifted.
This is a general feature of three-band magnon-phonon hybridization that one state is uncoupled to the magnon and remains a pure phonon, while the other two states are magnon polarons with equal proportions of magnons and phonons~\cite{supplement}.
Additionally, the hybridization generates phonon angular momentum for the states close to the avoided crossing [cf. \cref{fig:hybridized_system}(c)].
Interestingly, we find that the phonon angular momentum changes sign and vanishes directly at the avoided crossing.
Only in its vicinity it becomes nonzero.
Further analysis reveals that the sublattice-resolved phonon angular momentum is maximized directly at the avoided crossing reaching values of up to $\hbar / 2$~%
\footnote{%
    The saturation at fractions of $\hbar$ is a result of the normalization of the wave function, which, for the central pure phonon state, is normalized across the two sublattice degrees of freedom and for the hybridized states it is normalized across both the sublattice and magnon/phonon sector.
    Thus, the sublattice-resolved phonon angular momentum of the former is approximately $\hbar/2$ and that of the latter is about $\hbar/4$.%
}
and the lattice vibrations are almost fully circularly polarized [cf. \cref{fig:hybridized_system}(e), (f)].
Because the circular polarization is reversed on the two sublattices, the sublattice-resolved angular momenta partially compensate each other and the net angular momentum is strongly suppressed.

To analyze the cause of the compensation of the phonon angular momentum, we focus on the crossing of the lower magnon band with two phonon bands in \cref{fig:hybridized_system}(b).
\Cref{fig:coupling_strength} indicates that the lower magnon mode couples almost exclusively to phonons on the down-sublattice ($\downarrow$) and with a phonon angular momentum of $+\hbar$.
\Cref{fig:hybridized_system}(e) displays the phonon angular momentum of the $\downarrow$ sublattice.
Indeed, the magnon polaron bands exhibit a positive phonon angular momentum, while the uncoupled phonon mode has a negative phonon angular momentum.
At the (avoided) crossing the chirality-selective magnon-phonon coupling superimposes the linearly polarized phonons to right- and left-handed chiral phonons and splits them by selectively hybridizing the magnon and the phonon with matching handedness.
The uncoupled phonon consequently has opposite handedness.
On the other hand, the opposite phonon angular momentum on the $\uparrow$ sublattice can be explained by the phase differences between the sublattices of the crossing phonon bands (cf.~End Matter).

\begin{figure}
    \centering
    \includegraphics[width=\linewidth]{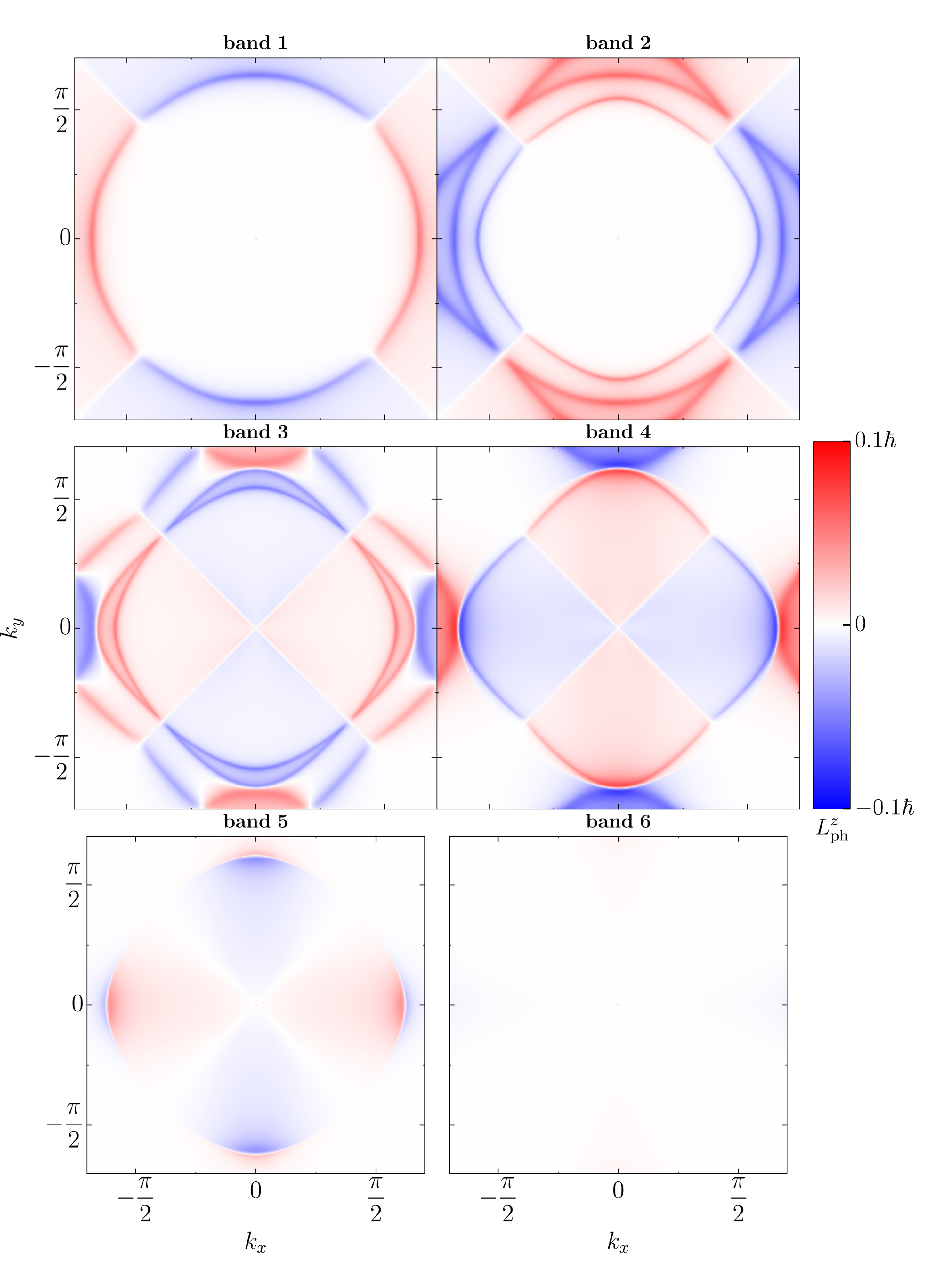}
    \caption{%
        Phonon angular momentum for all bands across the entire Brillouin zone. Nonzero angular momentum occurs around regions of avoided crossings and clearly reflects the $d$-wave character of the altermagnet.
        Bands are labeled in descending order starting with the highest energy.
        Parameters same as in \cref{fig:hybridized_system}.
    }
    \label{fig:density_plot_bz}
\end{figure}

Finally, \cref{fig:density_plot_bz} depicts the total phonon angular momentum of all bands across the entire Brillouin zone.
The $d$-wave structure of the phonon angular momentum texture can be clearly observed.
The sign of the angular momentum alternates in the Brillouin zone and the phonon angular momentum vanishes along the two nodal lines at $\overline{\Gamma \mathrm{M}}$.
One can distinguish two kinds of distributions: there are broad, diffuse features as well as sharp line-shaped \enquote{hotspots}.
The former can be mostly recognized in areas where the band either has magnon character (e.g., inner part of band 4) or where the band is close to another magnon band (e.g., outer area of band 2).
The pronounced line-shaped features originate from avoided crossings between magnon and phonon bands.
Since the energy gaps are relatively small, the lines have small thicknesses.
For regions where the band exhibits phonon character and is energetically separated from magnon bands, the phonon angular momentum is vanishingly small (e.g., band 6 or inner areas of band 1 and 2).

\paragraph{Discussion and conclusion}
The alternating sign of the phonon angular momentum across the Brillouin zone represents a qualitatively new form of chiral phonon behavior, distinct from previously studied uniform or valley-contrasting chiral phonons~\cite{zhangAngularMomentumPhonons2014,zhangChiralPhononsHighSymmetry2015,cohClassificationMaterialsPhonon2023}.
This momentum-dependent angular-momentum texture directly results from the interplay between the underlying $d$-wave magnetic order and the chiral selection rules of the relativistic magnon–phonon coupling.
In our analysis we have focused on a particular chirality-selective part of the interfacial DMI, but chirality-selective forms of magnon-phonon coupling can emerge more generally in the framework of rotationally invariant spin-lattice coupling~\cite{ruckriegel_angular_2020,mankovskyAngularMomentumTransfer2022,weisenhofer_rotationally_2023}.
The considered interfacial DMI may generate additional terms stemming from the distance dependence of $D(R)$ lifting the perfect chiral selectivity in a $\vect{k}$-dependent fashion.
While it renders the phonon angular momentum texture more complex, its general features remain robust as we demonstrate in the Supplemental Material~\cite{supplement}.
The discovery of $d$-wave chiral phonons lays the foundation for realizing phononic counterparts to electronic responses in altermagnets.

$d$-wave altermagnets are known to exhibit a spin-splitter effect~\cite{gonzalez-hernandez_efficient_2021}.
Based on the symmetry we can predict a \emph{phonon angular momentum splitter effect}, where a temperature gradient generates a flow of phonon angular momentum.
If the temperature gradient is applied along a high-symmetry spin-polarized direction, the phonon angular momentum current is parallel or antiparallel to the gradient [\cref{fig:graphical_abstract}(b)].
If the temperature gradient is applied along a nodal line, the phonon angular momentum current is transversal to the gradient [\cref{fig:graphical_abstract}(c)].
This nontrivial angular dependence distinguishes the splitter effect from the phonon angular momentum Hall effect proposed by Park and Yang~\cite{parkPhononAngularMomentum2020}.
Alternatively, the behavior under time reversal enables experimental differentiation of the two because the Hall effect remains unchanged and the splitter effect changes its sign upon reversal of the Néel order.
Furthermore, the splitter effect is proportional to the relaxation time and dominates in clean samples, whereas dirty samples favor the intrinsic Hall effect~\cite{nagaosa_anomalous_2010}.
Experimental studies have reported on measurements of phonon angular momentum currents via voltage generation in heavy-metal contacts~\cite{ohe_chirality-induced_2024}, delineating a potential route for observing the phonon angular momentum splitter effect.

The applications are not limited to the splitter effect.
Altermagnets form a net magnetization when subjected to strain, which is known as piezomagnetism~\cite{steward_dynamic_2023,mcclarty_landau_2024}.
Since in ionic crystals the phonon angular momentum is accompanied by an orbital magnetic moment~\cite{juraschek_orbital_2019,shabala_axial_2025}, phonon piezomagnetism contributes in altermagnets.
In addition, phononic analogs of spin filters and spin torques~\cite{bai_observation_2022,karube_observation_2022}, and the nonlinear thermal Edelstein effect~\cite{li_magnonic_2020,trama_nonlinear_2025} enrich the research landscape for ($d$-wave) chiral phonons.

A general challenge for measuring phononic analogs of electronic and magnonic responses in altermagnets lies in their apparent symmetry equivalence.
For instance, observing the phonon angular momentum splitter effect requires disentangling it from the concomitant magnon spin splitter effect~\cite{weissenhoferAtomisticSpinDynamics2024,sarkar_spin-split_2025}.
Crucially, however, the two phenomena originate from distinct mechanisms: the magnon spin texture is nonrelativistic, whereas the phonon angular momentum is inherently relativistic.
Consequently, the magnon spin texture is constrained by the spin point group (neglecting relativistic corrections), while the phonon angular momentum is fundamentally governed by the magnetic point group.
This implies that, in contrast to the robust magnon spin, the phonon angular momentum depends sensitively on the Néel vector orientation.
A sufficiently strong magnetic field can reorient the Néel vector away from its easy axis, thereby changing the plane of spin precession.
This, in turn, alters which spin components couple to the phonons and thus controls the resulting phonon angular momentum.
Within the model presented here, an in-plane Néel vector eliminates the phonon angular momentum because out-of-plane spin precession cannot generate out-of-plane atomic motion in a two-dimensional system.
A weaker magnetic field applied along the easy axis shifts the magnon bands and therefore generates $s$-wave components for the chiral phonons.
Hence, chiral phonons generated by magnetism offer great tunability compared to other approaches.
In addition, the sensitivity of the effect to interfacial DMI suggests possible routes for electrically or structurally tunable phononic devices, which may be relevant for phononics and hybrid spin–lattice information processing.
This highlights altermagnets as tunable platforms for realizing chiral phonons.

Our theory is valid for insulating altermagnets with sufficient relativistic magnetoelastic coupling.
Among two-dimensional $d$-wave materials, monolayer Cr$_2$Te$_2$O and Cr$_2$Se$_2$O~\cite{cui_efficient_2023} realize the symmetry of our spin Hamiltonian (space group $P4/mmm$), but rely on a substrate that induces interfacial DMI.
Another possibility are Janus altermagnets such as monolayer V$_2$SeTeO~\cite{zhu_multipiezo_2024}.
Here, the substitution of Se by Te in one layer breaks the out-of-plane mirror symmetry in the magnetic V layer and induces interfacial DMI intrinsically.
Our results can also be generalized to three dimensions, where many more candidate materials have been identified~\cite{bai_altermagnetism_2024}.
Although we have focused on a $d$-wave altermagnet, imprinting phonon angular momentum by magnon-phonon coupling with chiral selection rules can be applied to $g$- and $i$-wave magnets as well, where higher-wave phonon angular momentum textures are generated.

\begin{acknowledgments}
\paragraph{Acknowledgments.}
This work was funded by the German Research Foundation (DFG) as part of the German Excellence Strategy--EXC3112/1--533767171 (Center for Chiral Electronics) and through TRR 277-328545488 (Project No. B04), TRR 173-268565370 (Project No. B13), and Project No.~504261060 (Emmy Noether Programme).
A.M. acknowledges support by the Dynamics and Topology Center (TopDyn) funded by the State of Rhineland-Palatinate.
\end{acknowledgments}

\paragraph{Data availability.}
Upon reasonable request, the data associated with this article can be made available on Zenodo~\cite{zenodo}.

\bibliography{refs}

\onecolumngrid
\section*{End Matter}
\twocolumngrid

\setcounter{equation}{0}
\renewcommand{\theequation}{A\arabic{equation}}
\paragraph{Magnon Hamiltonian.}
The spin Hamiltonian reads
\begin{align}
\begin{aligned}
    \hat{H}_{\mathrm{m}}
    &=
    -\frac{J_{\mathrm{AFM}}}{2}
    \sum_{\langle i,j \rangle}
    \vect{S}_i \cdot \vect{S}_j
    \\
    &\quad
    -\frac{J_{\mathrm{FM}_1}}{2}
    \sum_{\langle\langle i,j \rangle\rangle_1}
    \vect{S}_i \cdot \vect{S}_j
    -\frac{J_{\mathrm{FM}_2}}{2}
    \sum_{\langle\langle i,j \rangle\rangle_2}
    \vect{S}_i \cdot \vect{S}_j
    \ .
    \label{eq:spin_hamil}
\end{aligned}
\end{align}
Here, $J_{\mathrm{AFM}} < 0$ and $J_{\mathrm{FM}_{1,2}} > 0$ describe the antiferromagnetic and ferromagnetic interaction, respectively.
$\langle i, j \rangle$ runs over all nearest neighbors, while $\langle \langle i, j \rangle \rangle_l$ runs over all next-nearest neighbors with bonds located within light gray ($l = 1$) or dark gray squares ($l = 2$) in \cref{fig:dispersion_relation_and_lattice}(a).
The splitting of the magnon bands is given by
\begin{align}
    \Delta E
    =
    2 \hbar^2 S \Bigl( J_{\mathrm{FM}_2} - J_{\mathrm{FM}_1} \Bigr)
    \Bigl[ \cos(\sqrt{2} k_y a) - \cos(\sqrt{2} k_x a) \Bigr]
    \ .
\end{align}

\setcounter{equation}{0}
\renewcommand{\theequation}{B\arabic{equation}}
\paragraph{Phonon Hamiltonian.}
The dynamics of the atoms is governed by
\begin{align}
\begin{split}
    \hat{H}_{\mathrm{ph}}
    &=
    \sum_i
    \frac{\vect{p}_i^2}{2 M}
    +
    \frac{1}{2}
    \sum_{\langle i,j \rangle}
    \qty[
        \qty(
            K_{\mathrm{L}}^{(1)}
            -
            K_{\mathrm{T}}
        )
        \qty(
            \vect{u}_{ij}
            \cdot
            \uvect{R}_{ij}
        )^2
        +
        K_{\mathrm{T}}
        \vect{u}_{ij}^2
    ]
    \\
    &\quad
    +
    \frac{K_{\mathrm{L}}^{(2)}}{2}
    \sum_{\langle \langle i,j \rangle\rangle}
    \qty(
        \vect{u}_{ij}
        \cdot
        \uvect{R}_{ij}
    )^2
    \ ,
\end{split}
\end{align}
where $\vect{u}_{ij} = \vect{u}_j - \vect{u}_i$ and $\vect{p}_i$ and $\vect{u}_i$ are the momentum and position operator of an atom at lattice site $i$, respectively, describing the dynamics in the $xy$ plane. $\uvect{R}_{ij}$ refers to the unit bond vector between atoms at lattice sites $i$ and $j$, while $M$ is the mass of the nuclei. 

\setcounter{equation}{0}
\renewcommand{\theequation}{C\arabic{equation}}
\paragraph{Compensation of the phonon angular momentum.}
Due to the magnetism-induced downfolding, the longitudinal acoustic crosses with the transversal optical phonon branch, for which the atoms in the unit cell oscillate in phase and out of phase, respectively.
Mathematically, the phases are given by $\exp{\iu (\vect{k} + \vect{G}) \cdot \Delta \vect{b}}$ where $\Delta \vect{b} = \frac{a}{\sqrt{2}} (1\ 1\ 0)^{\mathrm{T}}$ is the bond vector between the basis atoms and $\vect{G}$ is either zero for the acoustic branch, or, for an optical branch, an appropriate reciprocal lattice vector of the magnetic Bravais lattice.
By choosing an arbitrary $\vect{k}$-point along $\overline{\Gamma \mathrm{Y}}$, represented by $\vect{k} = (0\ k_y\ 0)^{\mathrm{T}}$ and by setting $\vect{G} = \left(0\ \frac{2\pi}{\sqrt{2} a}\ 0\right)^{\mathrm{T}}$, one finds $\vect{u}_\downarrow = \pm \e^{- \iu \Phi} \vect{u}_\uparrow$ for the acoustic and optical branches, where $\vect{u}_{p}$ is the wavefunction of basis atoms $p = {\uparrow}, {\downarrow}$ and $\Phi = - \frac{a}{\sqrt{2}} k_y$.
The longitudinal $\ket{l} = \left( 0\ 1\ 0\ \e^{\iu \Phi} \right)^{\mathrm{T}}$ and transverse modes $\ket{t} = \left( 1\ 0\ -\e^{\iu \Phi}\ 0 \right)^{\mathrm{T}}$ are linearly polarized along $y$ and $x$ direction, respectively.
Here, we represent the wave functions in the
$
    \mqty(
        u_{\downarrow}^x
        &
        u_{\downarrow}^y
        &
        u_{\uparrow}^x
        &
        u_{\uparrow}^y
    )
$
basis.
If we now construct right-handed circularly polarized phonons ($\langle L_{\mathrm{ph}}^z \rangle = +\hbar$) on the $\downarrow$ sublattice, the $\uparrow$ sublattice will be left-handed circularly polarized:
\begin{align}
    \ket{t} + \iu \ket{l}
    =
    \mqty(
        1 & \iu & -\e^{\iu \Phi} & \iu \e^{\iu\Phi}
    )^{\mathrm{T}}
    \ ,
\end{align}
and vice versa.
Hence, by selectively coupling a certain circularly polarized phonon mode of the $\downarrow$ sublattice to the magnon band, the opposite chirality is naturally imposed on the $\uparrow$ sublattice.
Since the magnon, depending on $\vect{k}$, couples to both sublattices, the phonons are not circular, but elliptical.
The closer the crossing is located at the edge of the Brillouin zone, the more the magnons specifically couple to one sublattice (in this case $\downarrow$) and the ellipticity is modified.

\end{document}


\title{
\textit{Supplemental Material}
\vspace{0.25cm}
\hrule
\vspace{0.25cm}
D-Wave Phonon Angular Momentum Texture in Altermagnets by Magnon-Phonon-Hybridization
}

\author{Hannah Bendin\orcidlink{0009-0006-2155-2834}}
\affiliation{Institute of Physics, Martin Luther University Halle-Wittenberg and Halle-Berlin-Regensburg Cluster of Excellence \enquote{Center for Chiral Electronics}, D-06120 Halle (Saale), Germany}
\affiliation{Institute of Physics, Otto von Guericke University Magdeburg, D-39106 Magdeburg, Germany}
\author{Alexander Mook\orcidlink{0000-0002-8599-9209}}
\affiliation{Institute of Solid State Theory, University of Münster, D-48149 Münster, Germany}
\affiliation{Institute of Physics, Johannes Gutenberg University Mainz, D-55128 Mainz, Germany}
\author{Ingrid Mertig\orcidlink{0000-0001-8488-0997}}
\affiliation{Institute of Physics, Martin Luther University Halle-Wittenberg and Halle-Berlin-Regensburg Cluster of Excellence \enquote{Center for Chiral Electronics}, D-06120 Halle (Saale), Germany}
\author{Robin R.~Neumann\orcidlink{0000-0002-9711-3479}}
\email[Correspondence email address: ]{rneumann@uni-mainz.de}
\affiliation{Institute of Solid State Theory, University of Münster, D-48149 Münster, Germany}
\affiliation{Institute of Physics, Johannes Gutenberg University Mainz, D-55128 Mainz, Germany}
\affiliation{Institute of Physics, Martin Luther University Halle-Wittenberg and Halle-Berlin-Regensburg Cluster of Excellence \enquote{Center for Chiral Electronics}, D-06120 Halle (Saale), Germany}

\date{\today}

\maketitle

\tableofcontents

\newpage

\section{Bosonization of magnons and phonons}
\subsection{Magnon Hamiltonian}
We derive the magnon Hamiltonian by mapping the spin operators onto bosonic operators using the truncated Holstein-Primakoff transformation~\cite{holsteinFieldDependenceIntrinsic1940}:
\begin{subequations}
\begin{align}
    \frac{\vect{S}_{\uparrow, i}}{\hbar}
    &\approx
    \qty(
        S
        -
        \adj{a}_{\uparrow, i} a_{\uparrow, i}
    )
    \unitvect{z}
    +
    \sqrt{S}
    \qty(
        \unitvect{e}_+ 
        \adj{a}_{\uparrow, i} 
        +
        \unitvect{e}_- 
        a_{\uparrow, i}
    )
    \ ,
    \\
    \frac{\vect{S}_{\downarrow, i}}{\hbar}
    &\approx
    \qty(
        \adj{a}_{\downarrow, i} a_{\downarrow, i}
        -
        S
    )
    \unitvect{z}
    -
    \sqrt{S}
    \qty(
        \unitvect{e}_- 
        \adj{a}_{\downarrow, i} 
        +
        \unitvect{e}_+ 
        a_{\downarrow, i}
    )
    \ ,
\end{align}
\end{subequations}
where $a_{\updownarrows, i}$ and $\adj{a}_{\updownarrows, i}$ are bosonic annihilation and creation operators for spin-flip excitations in the $i$-th unit cell of the sublattice $\uparrow$ or $\downarrow$.
We have introduced the complex unit vectors
$
    \unitvect{e}_{\pm}
    =
    \frac{
        \unitvect{x}
        \pm
        \iu \unitvect{y}
    }{
        \sqrt{2}
    }
$.

After Fourier transformation
\begin{align}
    a_{\updownarrows, \vect{k}}
    =
    \frac{1}{\sqrt{N}}
    \sum_{i}
    \e^{-\iu \vect{k} \cdot \vect{R}_{\updownarrows, i}}
    a_{\updownarrows, i}\ ,
    \qquad
    \adj{a}_{\updownarrows, \vect{k}}
    =
    \frac{1}{\sqrt{N}}
    \sum_{i}
    \e^{\iu \vect{k} \cdot \vect{R}_{\updownarrows, i}}
    \adj{a}_{\updownarrows, i}\ ,
    \label{eq: fourier transformation}
\end{align}
where $N$ denotes the total number of unit cells and $\unitvect{R}_{\uparrow, i}$ ($\unitvect{R}_{\downarrow, i}$) is the lattice vector of sublattice $\uparrow$ ($\downarrow$) in unit cell $i$, the bilinear part of the Hamiltonian can be written as
\begin{align}
    \hamil_{\mathrm{m}}
    &=
    \frac{1}{2}
    \sum_{\vect{k}}
    \adjvect{\alpha}_{\vect{k}}
    \hmatr_{\vect{k}}^{\mathrm{m}}
    \vect{\alpha}_{\vect{k}}
    \label{eq: bilinear Hamiltonian}
\end{align}
with $\adjvect{\alpha}_{\vect{k}} = (\adj{a}_{\uparrow,\vect{k}}\ \adj{a}_{\downarrow,\vect{k}}\ a_{\uparrow,-\vect{k}}\ a_{\downarrow,-\vect{k}})$. The kernel of the Hamiltonian in the main text reads
\begin{subequations}
\begin{align}
    \hmatr_{\vect{k}}^{\mathrm{m}} = \mqty(
        \xi_{\mathrm{AFM}}  &   &   &   \eta_{\vect{k}}   \\
            &   \xi_{\mathrm{AFM}}  &   \eta_{\vect{k}}   &   \\
            &   \eta_{\vect{k}}   &   \xi_{\mathrm{AFM}}  &   \\
        \eta_{\vect{k}}   &   &   &   \xi_{\mathrm{AFM}}
    )
    + \mqty(
        \xi_{\mathrm{FM}}^{(1)} &   &   &   \\
            &   \xi_{\mathrm{FM}}^{(2)}  &  &   \\
            &   &   \xi_{\mathrm{FM}}^{(1)} &   \\
            &   &   &\xi_{\mathrm{FM}}^{(2)}    \\
    )\ ,
\end{align}
where $\xi_{\mathrm{AFM}} = -4 \hbar^2 S J_{\mathrm{AFM}}$ and 
\begin{align}
\begin{aligned}
    \xi_{\mathrm{FM}}^{(1)} &=  2 \hbar^2 S \qty{ J_{\mathrm{FM}_1} \left[ 1 - \cos(\sqrt{2} k_x a) \right] + J_{\mathrm{FM}_2} \left[ 1 - \cos(\sqrt{2} k_y a) \right]  }\ ,   \\
    \xi_{\mathrm{FM}}^{(2)} &=  2 \hbar^2 S \qty{ J_{\mathrm{FM}_1} \left[ 1 - \cos(\sqrt{2} k_y a) \right] + J_{\mathrm{FM}_2} \left[ 1 - \cos(\sqrt{2} k_x a) \right] }\ , \\
    \eta_{\vect{k}}       &=  2 \hbar^2 S J_{\mathrm{AFM}} \left\{ \cos\left[ \frac{a}{\sqrt{2}} (k_x + k_y) \right] + \cos\left[ \frac{a}{\sqrt{2}} (k_x - k_y) \right] \right\}\ .
\end{aligned}
\end{align}
\label{eq:magnon_kernel}
\end{subequations}
To diagonalize the matrix, we apply a Bogoliubov transformation
\begin{align}
    \adj{\qty(\matr{T}_{\vect{k}}^{\mathrm{m}})}
    \hmatr_{\vect{k}}^{\mathrm{m}}
    \matr{T}_{\vect{k}}^{\mathrm{m}}
    =
    \matr{\mathcal{E}}_{\vect{k}}
    =
    \diag \qty(\mqty{
        \varepsilon_{\uparrow \vect{k}} & \varepsilon_{\downarrow \vect{k}}
        &
        \varepsilon_{\uparrow, -\vect{k}} & \varepsilon_{\downarrow, -\vect{k}}
    })
    \ ,
    \label{eq:bog_trafo}
\end{align}
which conserves the bosonic commutation relations by enforcing~\cite{colpaDiagonalizationQuadraticBoson1978,shindou_topological_2013}
\begin{align}
    \adj{\qty(\matr{T}_{\vect{k}}^{\mathrm{m}})}
    \metricmatr
    \matr{T}_{\vect{k}}^{\mathrm{m}}
    =
    \metricmatr
    \ .
\end{align}
Here, $\metricmatr = \diag\mqty(1 & 1 & -1 & -1)$ is the bosonic metric.
The algorithm for the transformation is detailed in Ref.\,\cite{colpaDiagonalizationQuadraticBoson1978}.

\subsection{Phonon Hamiltonian}
To treat both magnons and phonons on equal footing and facilitate the computation and analysis of hybridized states, the phonon Hamiltonian will also be written in second quantization. However, rather than quantizing in terms of normal modes, as is commonly done in standard textbooks, we adopt an approach of local harmonic oscillators that is described in more detail in Refs.\,\cite{neumannTheoreticalPredictionProbing2024,bissbortOperatorbasedDerivationPhonon2016}.

For this method, the general phonon Hamiltonian,
\begin{subequations}
\begin{align}
    \hat{H}_{\mathrm{ph}} = \sum_{i} \frac{\vect{p}_i^2}{2M} + \frac{1}{2} \sum_{i,j} \trpvect{u}_i \matr{\Phi}_{ij} \vect{u}_j\ ,
\end{align}
can be decomposed into two contributions
\begin{align}
    \hat{H}_{\mathrm{ph}} = 
        \underbrace{
            \sum_{i \vphantom{\substack{j\\j}}} \left[ \frac{\vect{p}_i^2}{2M} + \frac{1}{2} \trpvect{u}_i \matr{\Phi}_{ii}^{\mathrm{diag}} \vect{u}_i \right]
        }_{=\hat{H}_{\mathrm{loc}}}
        +
        \underbrace{
            \frac{1}{2} \sum_{i,j} \trpvect{u}_i \bigl[ \matr{\Phi}_{ij} - \delta_{ij} \matr{\Phi}_{ij}^{\mathrm{diag}} \bigr] \vect{u}_j
        }_{=\hat{H}_{\mathrm{int}}}\ ,
\end{align}
\end{subequations}
where $\matr{\Phi}_{ij}^{\mathrm{diag}}$ is the matrix containing only the diagonal elements of $\matr{\Phi}_{ij}$. The operator $\hat{H}_{\mathrm{loc}}$ describes a sum of independent one-dimensional harmonic oscillators that are localized at certain lattice sites $i$ and oscillate along single spatial directions. Their quantization is given by
\begin{subequations}
\begin{align}
    u_{\updownarrows \alpha i}
    =
    \sqrt{\frac{\hbar}{2M\Omega}}
    \bigl(
    b_{\updownarrows \alpha i}^{\dagger}
    +
    b_{\updownarrows \alpha i}
    \bigr)
    \ ,
    \qquad
    p_{\updownarrows \alpha i}
    =
    \iu
    \sqrt{\frac{\hbar M\Omega}{2}}
    \bigl(
    b_{\updownarrows \alpha i}^{\dagger}
    -
    b_{\updownarrows \alpha i}
    \bigr)
\end{align}
where 
\begin{align}
    \Omega = \sqrt{\frac{\Phi}{M}}
\end{align}
\end{subequations}
is the eigenfrequency of the local oscillators and $\alpha = x, y$ denotes the spatial direction (we focus on the two-dimensional case studied in the main text). Here, we exploited the fact that the elastic interactions $\Phi = \Phi_{ii}^{\diag}$ are identical in $x$ and $y$ direction and for both sublattices. The remaining terms, $\hat{H}_{\mathrm{int}}$, describe the interactions between the local harmonic oscillators. Thus, the approach is analogous to the local bosonization of magnons introduced by the Holstein-Primakoff transformation. 

Once again applying a Fourier transformation yields the matrix representation of the Hamiltonian,
\begin{align}
    \hamil_{\mathrm{ph}}
    &=
    \frac{1}{2}
    \sum_{\vect{k}}
    \adjvect{\beta}_{\vect{k}}
    \hmatr_{\vect{k}}^{\mathrm{ph}}
    \vect{\beta}_{\vect{k}}
    \ ,
    \label{eq:bilinear_phonon_hamil}
\end{align}
where $\adjvect{\beta}_{\vect{k}} = (\adj{b}_{\uparrow x \vect{k}}\ \adj{b}_{\uparrow y \vect{k}}\ \adj{b}_{\downarrow x \vect{k}}\ \adj{b}_{\downarrow y \vect{k}}\ b_{\uparrow x (-\vect{k})}\ b_{\uparrow y (-\vect{k})}\ b_{\downarrow x (-\vect{k})}\ b_{\downarrow y (-\vect{k})})$ and the matrix can be diagonalized using a Bogoliubov transformation.

For the model investigated in the main text, the phonon kernel reads
\begin{subequations}
\begin{align}
    \matr{H}_{\vect{k}}^{\mathrm{ph}}
    &=
    \hbar
    \begin{pmatrix}
        \matr{D}_{\vect{k}} + \varOmega \idmatr
        &
        \matr{D}_{\vect{k}} - \varOmega \idmatr
        \\
        \matr{D}_{\vect{k}} - \varOmega \idmatr
        &
        \matr{D}_{\vect{k}} + \varOmega \idmatr
    \end{pmatrix}
    \ ,
    \\
    \matr{D}_{\vect{k}}
    &=
    \matr{D}_{\vect{k}}^{\mathrm{(nn)}}
    +
    \matr{D}_{\vect{k}}^{\mathrm{(nnn)}}
    \ ,
    \\
    \matr{D}_{\vect{k}}^{\mathrm{(nn)}}
    &=
    \frac{2 K_{\mathrm{L}}^{(1)}}{M \varOmega}
    \begin{pmatrix}
        1 - \frac{1}{2} f_+(\vect{k})
        &
        -\frac{1}{2} f_-(\vect{k})
        \\
        -\frac{1}{2} f_-(\vect{k})
        &
        1 - \frac{1}{2} f_+(\vect{k})
    \end{pmatrix}
    +
    \frac{2 K_{\mathrm{T}}^{(1)}}{M \varOmega}
    \begin{pmatrix}
        1 - \frac{1}{2} f_+(\vect{k})
        &
        \frac{1}{2} f_-(\vect{k})
        \\
        \frac{1}{2} f_-(\vect{k})
        &
        1 - \frac{1}{2} f_+(\vect{k})
    \end{pmatrix}
    \ ,
    \\
    \matr{D}_{\vect{k}}^{\mathrm{(nnn)}}
    &=
    \frac{4 K_{\mathrm{L}}^{(2)}}{M \varOmega}
    \begin{pmatrix}
        \sin^2 \frac{k_x a}{\sqrt{2}} & 0
        \\
        0 & \sin^2 \frac{k_y a}{\sqrt{2}}
    \end{pmatrix}
    \ ,
\end{align}
\end{subequations}
where 
$
    f_{\pm}(\vect{k})
    =
    \cos[(k_x + k_y) a / \sqrt{2}]
    \pm
    \cos[(k_x - k_y) a / \sqrt{2}]
$.
and
$
    \varOmega
    =
    \sqrt{
        \frac{
            2 K_{\mathrm{L}}^{(1)}
            +
            2 K_{\mathrm{L}}^{(2)}
            +
            2 K_{\mathrm{T}}^{(1)}
        }{M}
    }
$.
We have denoted the ($2 \times 2$) identity matrix as $\idmatr$.

In the basis of second quantization, the phonon angular momentum operator [Eq.\,(1) of the main text] and the corresponding eigenmodes of left (L)- and right (R)-circularly polarized phonons are given by
\begin{align}
    L_{\mathrm{ph}}^z = \mathrm{i} \hbar \sum_{\vect{k}} \sum_{p = \uparrow,\downarrow} \bigl( b_{px\vect{k}} \adj{b}_{py\vect{k}} - \adj{b}_{px\vect{k}} b_{py\vect{k}} \bigr)\ ,
    \qquad
    \adj{b}_{\updownarrows R/L\vect{k}} = \frac{1}{\sqrt{2}} \bigl( \adj{b}_{\updownarrows x \vect{k}} \pm \mathrm{i} \adj{b}_{\updownarrows y \vect{k}} \bigr)\ .
\end{align}
We note in passing that we find a different sign in $L_{\text{ph}}^z$ compared to Romao \textit{et al.}~\cite{romao_chiral_2023}.
In the same spirit, we define the sublattice-resolved phonon angular momentum as
\begin{align}
    L_{\mathrm{ph},p}^z = \mathrm{i} \hbar \sum_{\vect{k}} \bigl( b_{px\vect{k}} \adj{b}_{py\vect{k}} - \adj{b}_{px\vect{k}} b_{py\vect{k}} \bigr)\ ,
    \label{eq:sublattice_pam}
\end{align}
which is connected to the net angular momentum as
$
    L_{\mathrm{ph}}^z
    =
    \sum_{p = \uparrow, \downarrow}
    L_{\mathrm{ph},p}^z
$.

\section{Magnon-phonon coupling}
\subsection{Interfacial Dzyaloshinskii-Moriya interaction}
Representing the in-plane Dzyaloshinskii-Moriya (DM) vector by $\vect{D}_{ij} = D \unitvect{R}_{ij} \times \unitvect{z}$, the Hamiltonian of the DM interaction is given by
\begin{align}
    \hat{H}_{\mathrm{DMI}} = \sum_{i,j}\vect{D}_{ij} \cdot \bigl(\vect{S}_i \times \vect{S}_j \bigr) = \sum_{i,j} D_{ij} \biggl[ (\vect{S}_i \cdot \unitvect{R}_{ij}) S_j^z - (\vect{S}_j \cdot \unitvect{R}_{ij}) S_i^z \biggr]\ .
    \label{eq:dmi}
\end{align}
This Hamiltonian can be expanded with respect to the (relative) displacements $\vect{u}_{ij} = \vect{R}_{ij} - \vect{R}_{ij}^{(0)}$ about the equilibrium positions/bonds $\vect{R}_{ij}^{(0)}$.
For the first order in $\vect{u}_{ij}$, we consider the spatial derivative for a particular dimer of spins at sites $i$ and $j$ (the indices of $R$ and $D$ are omitted for improved readability):
\begin{align}
    \begin{split}
        \pdv{}{R_{\alpha}}
        \qty[
            \vect{D}_{ij}
            \cdot
            \qty(
                \vect{S}_i
                \times
                \vect{S}_j
            )
        ]
        &=
        \pdv{}{R_{\alpha}}
        \qty[
            \frac{D}{R}
            \qty(
                \vect{S}_i
                S_j^z
                -
                \vect{S}_j
                S_i^z
            )
            \cdot
            \vect{R}_{ij}
        ]
        \\
        &=
        \sum_{\beta}
        \qty(
            S_i^\beta
            S_j^z
            -
            S_j^\beta
            S_i^z
        )
        \pdv{}{R_{\alpha}}
        \qty(
            \frac{D}{R}
            R^\beta
        )
        \\
        &=
        \sum_{\beta}
        \qty(
            S_i^\beta
            S_j^z
            -
            S_j^\beta
            S_i^z
        )
        \qty(
            \frac{D}{R}
            \kron{\alpha}{\beta}
            +
            R^\beta
            \pdv{}{R^\alpha}
            \frac{D}{R}
        )
        \ ,
        \label{eq: DMI derivative}
    \end{split} 
\end{align}
which can be utilized for a Taylor expansion of \cref{eq:dmi} up to first order. The first term in the last expression accounts for a change of the \emph{direction} of the DM vector by bond rotations, while the latter term, which can be expressed as
\begin{align}
    R^{\beta} \pdv{}{R^{\alpha}} \frac{D}{R} = \frac{D}{R} \left( 1 - \frac{R}{D} \pdv{D}{R} \right) \hat{R}^{\alpha} \hat{R}^{\beta}\ ,
\end{align}
incorporates the effects of changes of the interaction \emph{strength} -- explicitly via $\nicefrac{1}{R}$ as well as implicitly due to $D = D(R)$ -- due to modulations of the bond length.
For the calculations in the main manuscript, only the former contributions,
\begin{align}
    \hat{H}_{\mathrm{DMI}}^{(1)} = \sum_{i,j} \frac{D}{R} \left( S_i^{\alpha} S_j^z - S_j^{\alpha} S_i^z \right) \left( u_i^{\alpha} - u_j^{\alpha} \right)\ ,
\end{align}
have been taken into account.
Expanding the above expression in $\nicefrac{1}{S}$, we substitute $S_i^z$ by its mean field $\hbar S z_i$, where $z_i = +1$ for sublattice $\uparrow$ and $z_i = -1$ for sublattice $\downarrow$, and we obtain
\begin{align}
    \hat{H}_{\mathrm{DMI}}^{(1)}
    =
    \hbar S
    \sum_{i,j}
    \sum_{\alpha}
    \frac{D}{R} \left( S_i^{\alpha} z_j - S_j^{\alpha} z_i \right) \left( u_i^{\alpha} - u_j^{\alpha} \right)\ .
\end{align}
For next-nearest neighbor interactions, $z_j = z_i = \pm 1$ holds, where the sign is determined by the sublattice of the two interacting spins in agreement with Eq.~(2) of the main text.

We solve for the eigenstates of the coupled system by applying the same quantization procedures for magnons and phonons outlined above and expanding the term $\hat{H}_{\mathrm{DMI}}$ up to first order in the bosonic magnon operators, which yields the bilinear coupling Hamiltonian of the form
\begin{align}
    \mathcal{H}_{\mathrm{mpc}}
    =
    \frac{1}{2}
    \sum_{\vect{k}}
    \adjvect{\alpha}_{\vect{k}} \matr{H}_{\vect{k}}^{\mathrm{mpc}} \vect{\beta}_{\vect{k}}
    +
    \hc
\end{align}
with the block matrix structure 
\begin{align}
    \hmatr_{\vect{k}}^{\mathrm{mpc}}
    &=
    \begin{pmatrix}
        \matr{A}_{\vect{k}}^{\mathrm{mpc}} & \matr{A}_{\vect{k}}^{\mathrm{mpc}}
        \\
        \conj{\qty(\matr{A}_{-\vect{k}}^{\mathrm{mpc}})} & \conj{\qty(\matr{A}_{-\vect{k}}^{\mathrm{mpc}})}
    \end{pmatrix}
\end{align}
of its kernel.
For the spin-lattice coupling used in the main text [Eq.~(2)], we obtain
\begin{align}
    \matr{A}_{\vect{k}}^{\mathrm{mpc}}
    &=
    \frac{2 D}{a}
    \qty[
        2
        -
        \cos(\sqrt{2} k_x a)
        -
        \cos(\sqrt{2} k_y a)
    ]
    \sqrt{\frac{S^3}{2 M \varOmega}}
    \begin{pmatrix}
        1 & \iu & 0 & 0
        \\
        0 & 0 & 1 & -\iu
    \end{pmatrix}
    \ .
    \label{eq:mpc_kernel}
\end{align}
The complete matrix representation of the full coupled magnon-phonon system is given by
\begin{align}
    \matr{H}_{\vect{k}} = \begin{pmatrix}
        \matr{H}_{\vect{k}}^{\mathrm{m}} &   \matr{H}_{\vect{k}}^{\mathrm{mpc}}   \\
        (\matr{H}_{\vect{k}}^{\mathrm{mpc}})^{\dagger}   &   \matr{H}_{\vect{k}}^{\mathrm{ph}}
    \end{pmatrix}\ ,
\end{align}
where $\matr{H}_{\vect{k}}^{\mathrm{m}}$, $\matr{H}_{\vect{k}}^{\mathrm{ph}}$ and $\matr{H}_{\vect{k}}^{\mathrm{mpc}}$ are the matrix representations of the magnon, phonon and spin-lattice coupling Hamiltonian, respectively.

\subsection{Characterization of the chiral selectivity of the magnon-phonon coupling}
\label{sec:mpc_select}
To assess the chiral selectivity of the magnon-phonon coupling, we project $\hmatr_{\vect{k}}^{\mathrm{mpc}}$, which is defined in the basis of the Holstein-Primakoff modes in the magnon sector and the linearly polarized sublattice-resolved elastic modes in the phonon sector, onto the normal modes of the spin Hamiltonian and the circularly polarized sublattice-resolved elastic modes in the phonon sector.
The normal modes $\adj{\tilde{a}}_{\lambda \vect{k}}$ and $\tilde{a}_{\lambda \vect{k}}$ of the spin Hamiltonian, comprised by the Nambu spinors $\adj{\tilde{\vect{\alpha}}}_{\vect{k}}$ and $\tilde{\vect{\alpha}}_{\vect{k}}$, are generated by [cf.~\cref{eq:bog_trafo}]
\begin{align}
    \adjvect{\alpha}_{\vect{k}}
    =
    \adj{\tilde{\vect{\alpha}}}_{\vect{k}}
    \adj{\qty(\matr{T}_{\vect{k}}^{\mathrm{m}})}
    \ .
\end{align}
The transformation of the phonon operators from the Cartesian $\adjvect{\beta}_{\vect{k}}$ [cf.~\cref{eq:bilinear_phonon_hamil}] to the chiral basis
\begin{align}
    \adjvect{\beta}_{\mathrm{cp}, \vect{k}}
    =
    \mqty(
        \adj{\beta}_{\uparrow R \vect{k}}
        &
        \adj{\beta}_{\uparrow L \vect{k}}
        &
        \adj{\beta}_{\downarrow R \vect{k}}
        &
        \adj{\beta}_{\downarrow L \vect{k}}
        &
        \beta_{\uparrow R (-\vect{k})}
        &
        \beta_{\uparrow L (-\vect{k})}
        &
        \beta_{\downarrow R (-\vect{k})}
        &
        \beta_{\downarrow L (-\vect{k})}
    )
\end{align}
is mediated by a $\vect{k}$-independent matrix:
\begin{subequations}
\begin{align}
    \adjvect{\beta}_{\vect{k}}
    &=
    \adjvect{\beta}_{\mathrm{cp}, \vect{k}}
    \adjmatr{T}_{\mathrm{cp}}
    \ ,
    \\
    \adjmatr{T}_{\mathrm{cp}}
    &=
    \frac{1}{\sqrt{2}}
    \begin{pmatrix}
        1 & -\iu & 0 & 0 & 0 & 0 & 0 & 0
        \\
        1 & \iu & 0 & 0 & 0 & 0 & 0 & 0
        \\
        0 & 0 & 1 & -\iu & 0 & 0 & 0 & 0
        \\
        0 & 0 & 1 & \iu & 0 & 0 & 0 & 0
        \\
        0 & 0 & 0 & 0 & 1 & \iu & 0 & 0
        \\
        0 & 0 & 0 & 0 & 1 & -\iu & 0 & 0
        \\
        0 & 0 & 0 & 0 & 0 & 0 & 1 & \iu
        \\
        0 & 0 & 0 & 0 & 0 & 0 & 1 & -\iu
    \end{pmatrix}
    \ .
\end{align}
\end{subequations}
The chiral phonon operators $\beta_{p l \vect{k}}$ and $\adj{\beta}_{p l \vect{k}}$ with $p = \uparrow, \downarrow$ and $l = R, L$ are defined such that the sublattice-resolved phonon angular momenta $L_{\mathrm{ph}, p}^z$ [\cref{eq:sublattice_pam}] become diagonal when expressed in that basis.
Combining the two transformations, we can recast the spin-lattice coupling matrix in the form
\begin{align}
    \hamil_{\mathrm{mpc}}
    &=
    \frac{1}{2}
    \sum_{\vect{k}}
    \adj{\tilde{\vect{\alpha}}}_{\vect{k}}
    \tilde{\hmatr}_{\vect{k}}^{\mathrm{mpc}}
    \vect{\beta}_{\mathrm{cp}, \vect{k}}
    +
    \hc
    \ ,
    \\
    \tilde{\hmatr}_{\vect{k}}^{\mathrm{mpc}}
    &=
    \adj{\qty(\matr{T}_{\vect{k}}^{\mathrm{m}})}
    \hmatr_{\vect{k}}^{\mathrm{mpc}}
    \matr{T}_{\mathrm{cp}}
    =
    \begin{pmatrix}
        \tilde{\matr{A}}_{\vect{k}}^{\mathrm{mpc}}
        &
        \tilde{\matr{A}}_{\vect{k}}^{\mathrm{mpc}}
        \\
        \conj{\qty(\tilde{\matr{A}}_{-\vect{k}}^{\mathrm{mpc}})}
        &
        \conj{\qty(\tilde{\matr{A}}_{-\vect{k}}^{\mathrm{mpc}})}
    \end{pmatrix}
    \ .
\end{align}
We can now quantify the chiral selectivity of the coupling: the coupling of the $\lambda$-th magnon band to the phonon localized on sublattice $p$ with the chirality $l$ is given by
$
    \abs{\tilde{\matr{A}}_{\vect{k}}^{\mathrm{mpc}}}_{\lambda, p l}
$.
This is the quantity we have plotted in Fig.~3 of the main text as well as in \cref{fig:coupling_strength nnn G=0} and \cref{fig:coupling_strength nn G=0}

\subsection{Characterization of magnon polaron eigenmodes}
Here, we describe how we determine the magnon and phonon character of a magnon polaron eigenmode, by decomposing its wave function.
This forms the foundation for the coloring of the bands in Fig.~4(a) and (b) in main text.

By a mere rearrangement of the matrix elements, we can write the full Hamiltonian as
\begin{align}
    H
    &=
    \frac{1}{2}
    \sum_{\vect{k}}
    \adjvect{\phi}_{\vect{k}}
    \hmatr_{\vect{k}}
    \vect{\phi}_{\vect{k}}
    \ ,
\end{align}
where the Nambu spinor
\begin{align}
    \adjvect{\phi}_{\vect{k}}
    &=
    \mqty(
        \adj{a}_{\uparrow \vect{k}}
        &
        \adj{a}_{\downarrow \vect{k}}
        &
        \adj{b}_{\uparrow x \vect{k}}
        &
        \adj{b}_{\uparrow y \vect{k}}
        &
        \adj{b}_{\downarrow x \vect{k}}
        &
        \adj{b}_{\downarrow y \vect{k}}
        &
        a_{\uparrow (-\vect{k})}
        &
        a_{\downarrow (-\vect{k})}
        &
        b_{\uparrow x (-\vect{k})}
        &
        b_{\uparrow y (-\vect{k})}
        &
        b_{\downarrow x (-\vect{k})}
        &
        b_{\downarrow y (-\vect{k})}
    )
\end{align}
contains Holstein-Primakoff bosons $a_{m \vect{k}}$ and $\adj{a}_{m \vect{k}}$ and linear phononic bosons $b_{m \mu \vect{k}}$ and $\adj{b}_{m \mu \vect{k}}$.
$m = \uparrow, \downarrow$ is the sublattice index and $\mu = x, y$ is the Cartesian index representing the direction of linear polarization.
After the Bogoliubov diagonalization, the normal modes $c_{\lambda \vect{k}}$ and $\adj{c}_{\lambda \vect{k}}$ ($\lambda$ band index) are superpositions of spin and elastic excitations:
\begin{align}
    \adjvect{\psi}_{\vect{k}}
    =
    \mqty(
        \adj{c}_{1 \vect{k}}
        &
        \cdots
        &
        \adj{c}_{6 \vect{k}}
        &
        c_{1 (-\vect{k})}
        &
        \cdots
        &
        c_{6 (-\vect{k})}
    )
    =
    \adjvect{\phi}_{\vect{k}}
    \inv{\qty(\adjmatr{T}_{\vect{k}})}
    \ .
\end{align}
To compute the magnon character $P_{\lambda \vect{k}}^{(\mathrm{m})}$ of band $\lambda$ and wave vector $\vect{k}$, we (i) compute the number of magnon excitations
\begin{align}
    N_{p, \lambda \vect{k}}^{(\mathrm{m})}
    =
    \sum_{\vect{k}'}
    \qty[
        \expval{
            \adj{a}_{p \vect{k}'}
            a_{p \vect{k}'}
        }{
            \lambda \vect{k}
        }
        -
        \expval{
            \adj{a}_{p \vect{k}'}
            a_{p \vect{k}'}
        }{
            0
        }
    ]
    =
    \abs{\qty({T}_{\vect{k}})_{p \lambda}}^2
    +
    \abs{\qty({T}_{\vect{k}})_{p + \nbands, \lambda}}^2
\end{align}
on a given sublattice $p$ for a given mode $\ket{\lambda \vect{k}}$, where $\nbands = 6$ is the number of bands, (ii) sum over all sublattices $p$, and (iii) divide by the total number of bosons (magnons and phonons):
\begin{align}
    P_{\lambda \vect{k}}^{\mathrm{(m)}}
    =
    \frac{
        \sum_{p}
        \qty[
            \abs{\qty({T}_{\vect{k}})_{p \lambda}}^2
            +
            \abs{\qty({T}_{\vect{k}})_{p + \nbands, \lambda}}^2
        ]
    }{
        \sum_{j = 1}^{2 \nbands}
        \abs{\qty({T}_{\vect{k}})_{j \lambda}}^2
    }
    \ .
\end{align}
The sum over $p$ in the numerator iterates over the magnon sector including all sublattices (here, $p = 1, 2$), while the sum over $j = 1, \dots, 2 \nbands$ in the denominator iterates over both the magnon sector including all sublattices and the phonon sector including all sublattices and vibration dimensions.
$P_{\lambda \vect{k}}^{\mathrm{(m)}}$ is bounded between 0 and 1, where $P_{\lambda \vect{k}}^{\mathrm{(m)}} = 1$ represents a pure magnon state and $P_{\lambda \vect{k}}^{\mathrm{(m)}} = 0$ characterizes a pure phonon state.

\section{Hybridization at three-band crossing}
This section focuses on the hybridization at the three-band crossing, which we highlight in the main text.
We explain how the crossing is lifted and how the phonon angular momentum emerges.
Without magnon-phonon coupling, there is a degeneracy at which the magnon $\ket{m}$, longitudinal phonon $\ket{l}$ and transverse phonon state $\ket{t}$ have the same energy $E$.
At this crossing, we can project the Hamiltonian to the degenerate sub-Hilbert space.
In the most general case, it reads
\begin{align}
    \hmatr
    &=
    \begin{pmatrix}
        E & \eta & \xi
        \\
        \conj{\eta} & E & 0
        \\
        \conj{\xi} & 0 & E
    \end{pmatrix}
    ,
\end{align}
written in the $(\ket{m}, \ket{l}, \ket{t})$ basis.
$\eta$ and $\xi$ are the magnetoelastic coupling constants coupling $\ket{m}$ to $\ket{l}$ and $\ket{m}$ to $\ket{t}$, respectively.
Since magnetoelastic coupling does not couple phonon modes directly, there is no coupling between $\ket{l}$ and $\ket{t}$.
We find that
$
    \vect{v}
    =
    \frac{1}{\sqrt{\abs{\xi}^2 + \abs{\eta}^2}}
    \trp{\mqty(
        0
        &
        -\xi
        &
        \eta
    )}
$
is an eigenvector of $\hmatr$ with the eigenvalue $E$.
Therefore, the magnetoelastic coupling parameterized by $\eta$ and $\xi$ promotes an eigenstate, which is a pure phonon and unchanged in energy.
Choosing the orthogonal phonon state
$
    \vect{v}'
    =
    \frac{1}{\sqrt{\abs{\xi}^2 + \abs{\eta}^2}}
    \trp{\mqty(
        0
        &
        \conj{\eta}
        &
        \conj{\xi}
    )}
$,
we can rewrite $\hmatr$ in the $(\ket{m}, \ket{v'}, \ket{v}$) basis:
\begin{align}
    \tilde{\hmatr}
    =
    \begin{pmatrix}
        E & \sqrt{\abs{\eta}^2 + \abs{\xi}^2} & 0
        \\
        \sqrt{\abs{\eta}^2 + \abs{\xi}^2} & E & 0
        \\
        0 & 0 & E
    \end{pmatrix}
    .
\end{align}
Here, we see that the magnetoelastic coupling hybridizes $\ket{m}$ and $\ket{v'}$, while $\ket{v}$ is uncoupled to both $\ket{m}$ and $\ket{v'}$.
We have made use of the SU(2) degree of freedom in the degenerate sub-Hilbert space, which allows one to form arbitrary superpositions of the longitudinal and transverse phonons.
This argument holds generally and does not rely on the concrete form of the magnetoelastic coupling.
Hence, we can always find a pair of basis phonons states $\ket{v}$ and $\ket{v'}$, where $\ket{v}$ is uncoupled to $\ket{m}$ and $\ket{v'}$.
The coupled phonon state $\ket{v'}$ hybridizes with the magnon giving rise to two magnon polaron states
$
    \ket{mp, \pm}
    =
    \qty(
        \ket{m}
        \pm
        \ket{v'}
    )
    /
    \sqrt{2}
$
with energies $E_\pm = E \pm \sqrt{\abs{\eta}^2 + \abs{\xi}^2}$, i.e., the magnon polarons consist of equal superpositions of $\ket{m}$ and $\ket{v'}$.

\begin{figure}
    \centering
    \includegraphics[width=0.7\linewidth]{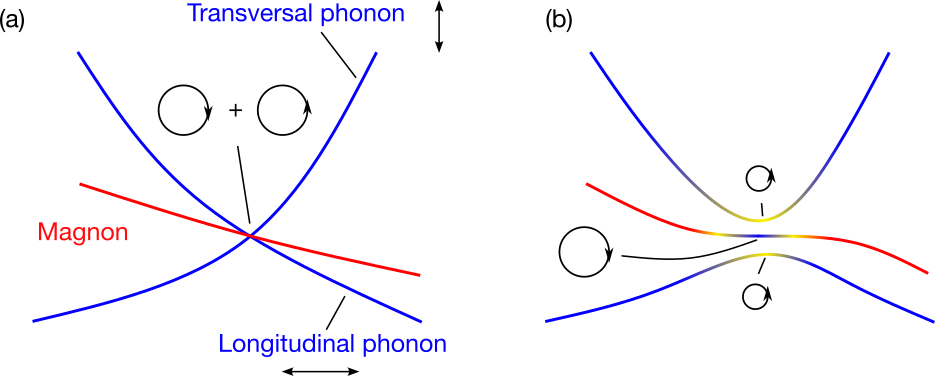}
    \caption{%
        Schematic picture of a crossing of two phonon bands and a magnon band.
        (a) Before hybridization the phonon bands have orthogonal linear polarizations.
        The straight double-arrows and the loop arrows indicate the phonon polarization of the relevant sublattice.
        At the avoided crossing, the eigenstates can be recombined to left- and right-handed circularly polarized modes.
        (b) The magnon hybridizes with its preferred circular polarization, while the remaining state is a pure phonon with opposite circular polarization.
    }
    \label{fig:3-crossing}
\end{figure}
Focusing on the concrete model of the paper, it turns out that if we choose a circular basis, only one of the phonon states couples with the magnon band, while the other one is uncoupled.
More precisely, if we consider the three-band crossing highlighted in Fig.~4(b) of the main text, $\ket{m}$ corresponds to the lower magnon band with spin $+\hbar$, $\ket{l}$ corresponds to the longitudinal acoustic phonon, and $\ket{t}$ to the transverse optical phonon [cf.~\cref{fig:3-crossing}(a)].
The hybridization leads to the mixing of all intersecting bands, which is why chiral quasiparticles emerge out of achiral ones.
The mixing of the phonons at the crossing due to magnon-phonon coupling creates a coupled phonon $\ket{v'}$, which is a right-handed circular phonon with $\expval{L_{\mathrm{ph}}^z} > 0$, and an uncoupled chiral phonon $\ket{v}$, which is a left-handed circular phonon with $\expval{L_{\mathrm{ph}}^z} < 0$ (w.r.t the sublattice $\downarrow$).
Therefore, two magnon polarons consisting of the magnon and the phonon of preferential chirality are formed and the phonon of opposite chirality remains at the same energy [cf.~\cref{fig:3-crossing}(b)].

\section{Additional results}
In the main text, we invoke some simplifications of the model.
Although the omitted interactions are crucial for describing realistic materials, they do not fundamentally change the core physics of this paper: the generation of $d$-wave chiral phonons by magnon-phonon hybridization.
Nevertheless, to demonstrate the robustness of our results, we lift some of the simplifications and consider modifications of our model in three parts: the phonon Hamiltonian, the magnon Hamiltonian, and the spin-lattice coupling.
As we demonstrate below, in all considered cases, the $d$-wave angular momentum texture remains robust and most changes are quantitative in nature.

\subsection{Impact of checkerboard elastic constants}
For the results in the main text we have assumed that the phonons can be described on a square lattice with only one atom per unit cell, whereas for the magnons we assumed Néel order and a checkerboard pattern that manifests itself in the anisotropic next-nearest neighbor Heisenberg coupling.
Naturally, such a symmetry breaking does not arise in a simple square lattice such that one could expect checkerboard pattern to appear in the phononic Hamiltonian as well.
In this section, we study the effect of such a checkerboard pattern in the phonon Hamiltonian.
We consider
\begin{align}
    \hat{H}_{\mathrm{ph}}
    &=
    \sum_{i}
    \frac{\vect{p}_i}{2 M}
    +
    \frac{1}{2}
    \sum_{\langle i, j \rangle}
    \qty[
        \qty(
            K_{\mathrm{L}}^{(1)}
            -
            K_{\mathrm{T}}
        )
        \qty(
            \vect{u}_{ij}
            \cdot
            \unitvect{R}_{ij}
        )^2
        +
        K_{\mathrm{T}}
        \vect{u}_{ij}^2
    ]
    +
    \frac{K_{\mathrm{L}, 1}^{(2)}}{2}
    \sum_{\langle\langle i, j \rangle\rangle_1}
    \qty(
        \vect{u}_{ij}
        \cdot
        \unitvect{R}_{ij}
    )^2
    +
    \frac{K_{\mathrm{L}, 2}^{(2)}}{2}
    \sum_{\langle\langle i, j \rangle\rangle_2}
    \qty(
        \vect{u}_{ij}
        \cdot
        \unitvect{R}_{ij}
    )^2
    ,
\end{align}
where the first sum runs over all nearest neighbor $\langle i, j \rangle$ and comprises longitudinal $K_{\mathrm{L}}^{(1)}$ and transversal restoring forces $K_{\mathrm{T}}$ and the second and third sums run over all next-nearest neighbors $\langle\langle i, j \rangle\rangle_l$ with bonds located within light gray ($l = 1$) or dark gray squares ($l = 2$) in Fig.~2(a), realizing anisotropic longitudinal restoring forces $K_{\mathrm{L}, 1}^{(2)}$ and $K_{\mathrm{L}, 2}^{(2)}$.

\begin{figure}
    \centering
    \includegraphics[width=0.5\linewidth]{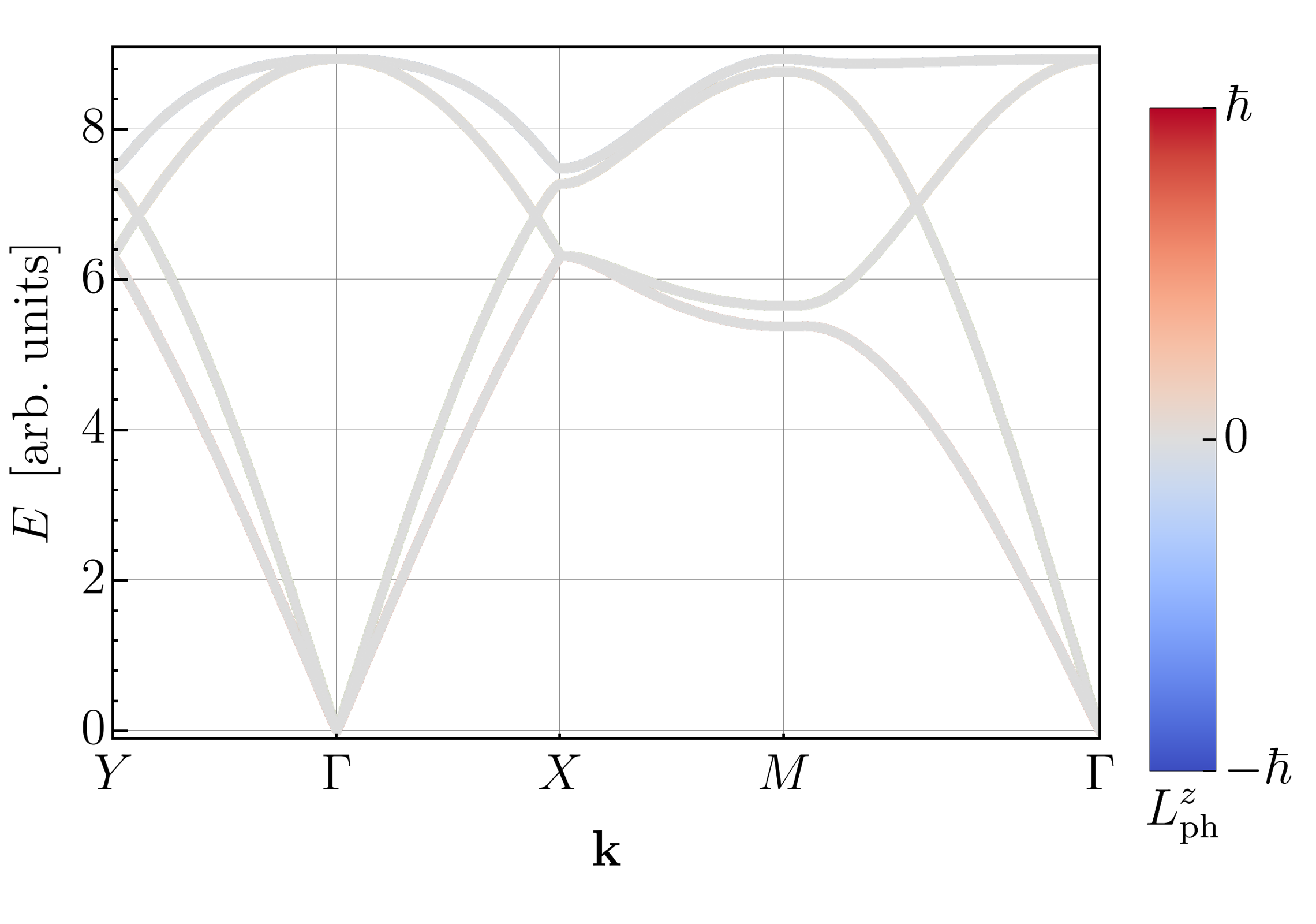}
    \caption{%
        Phonon band structure accounting for the checkerboard symmetry.
        The gray color indicates zero phonon angular momentum in the absence of magnon-phonon coupling.
        The parameters used for the calculation are
        $M = 10$,
        $K_{\mathrm{L}}^{(1)} = 160$,
        $K_{\mathrm{T}} = K_{\mathrm{L},2}^{(2)} = 40$,
        $K_{\mathrm{L},1}^{(2)} = 32.4$,
        and
        $a = 1$.
    }
    \label{fig:phonon_lz}
\end{figure}
The phonon band structure, visualized in \cref{fig:phonon_lz}, differs only at the edge of the first Brillouin zone, where a gap at the X and Y points opens between the upper band pair and the degeneracies of all bands are lifted along the high-symmetry path $\overline{\mathrm{XM}}$.
Importantly, the crossing between the acoustic longitudinal phonon band and the optical transverse phonon band remains stable.
Because the lattice has $PT$ symmetry, the phonon angular momentum vanishes despite the checkerboard pattern.

\begin{figure}
    \includegraphics[width=\textwidth]{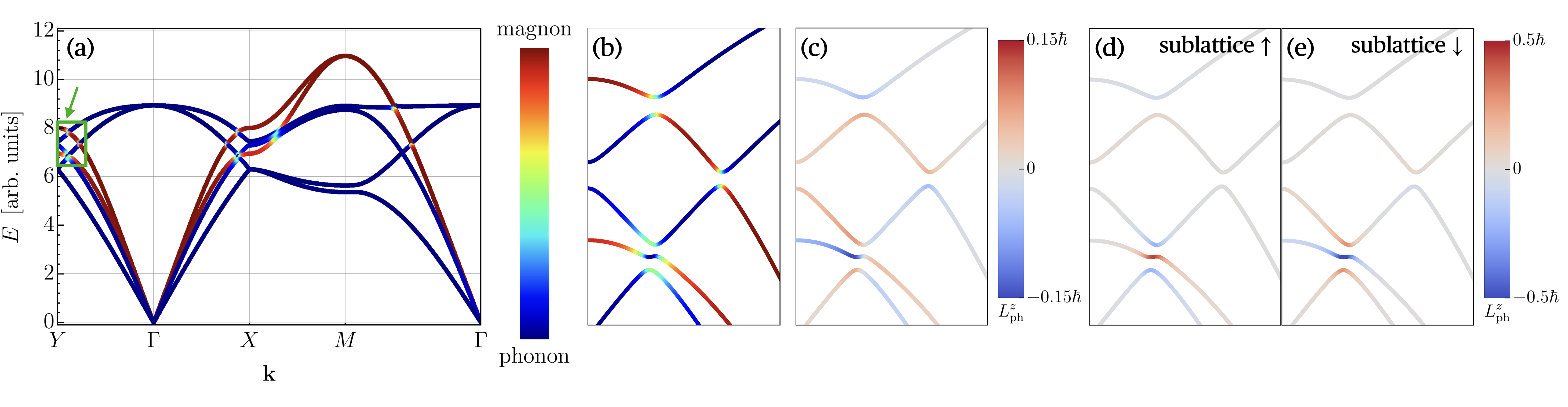}
    \caption{%
        Coupled magnon-phonon hybrid system accounting for the checkerboard symmetry in the bare phonon Hamiltonian.
        (a) Band structure of the magnon polarons along a high-symmetry path in the magnetic Brillouin zone.
        The color indicates the magnon/phonon character of the quasiparticles.
        The region highlighted by the green box and arrow is displayed in greater detail in (c)--(f).
        (b) Zoom of the avoided crossing pattern and quasiparticle character indicated by color.
        (c) Illustration of the same region but depicting the phonon angular momentum.
        (d),~(e) Sublattice-resolved phonon angular momentum for sublattice $\uparrow$ and sublattice $\downarrow$, respectively.
        The parameters used for the calculations are
        $J_{\mathrm{AFM}} = -1$,
        $J_{\mathrm{FM}_1} = 0.371$,
        $J_{\mathrm{FM}_2} = 0.5$,
        $S = 1$,
        $M = 10$,
        $K_{\mathrm{L}}^{(1)} = 160$,
        $K_{\mathrm{T}} = K_{\mathrm{L},2}^{(2)} = 40$,
        $K_{\mathrm{L},1}^{(2)} = 32.4$,
        and
        $a = 1$.
    }
    \label{fig:checkerboard phonon hamiltonian}
\end{figure}
Now we turn to the coupled system.
The magnon Hamiltonian and the magnon-phonon Hamiltonian are taken to be the same as in the main text.
Avoided crossings form where magnon bands and phonon bands would intersect [cf.~\cref{fig:checkerboard phonon hamiltonian}(a)].
At the point where the magnon band meets the two phonon band, two magnon polaron states and one pure phonon state emerge [\cref{fig:checkerboard phonon hamiltonian}(b)] and a net angular momentum emerges around all avoided crossings [\cref{fig:checkerboard phonon hamiltonian}(c)].
For the double-crossing, we find again that the sign of the net angular momentum changes and becomes zero at the avoided crossing because the sublattice-resolved phonon angular momenta cancel [\cref{fig:checkerboard phonon hamiltonian}(d),\,(e)].

\begin{figure}
    \includegraphics[width=\textwidth]{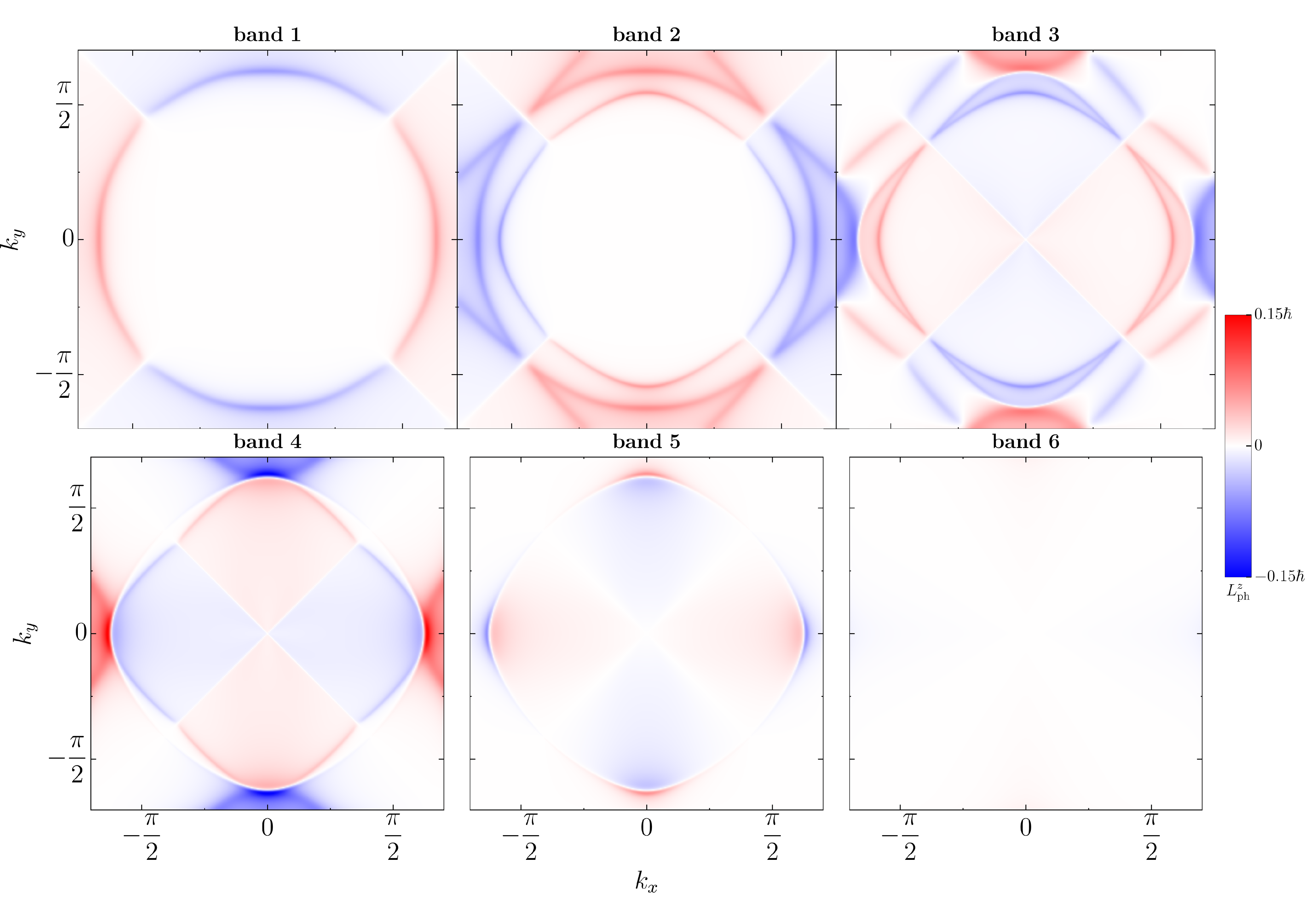}
    \caption{%
        Phonon angular momentum texture for all bands across the entire first Brillouin zone accounting for the checkerboard symmetry in the phonon Hamiltonian.
        Nonzero angular momentum occurs around regions of avoided crossings and clearly reflects the $d$-wave character of the altermagnet, thereby proving that the main result of our manuscript remains robust.
        Bands are labeled in descending order starting with the highest energy.
        Parameters same as in \cref{fig:checkerboard phonon hamiltonian}.%
    }
    \label{fig:checkerboard phonon lz bz}
\end{figure}
As a final verification, we plot the phonon angular momentum texture in \cref{fig:checkerboard phonon lz bz}.
Both the qualitative symmetries and the shape of the texture remains unchanged.
Compared to the calculation in the main text, the phonon angular momentum is of greater magnitude.
Overall, the checkerboard pattern in the phonon Hamiltonian has no qualitative effect on the phonon angular momentum texture, which is why we omit it in the main text.

\subsection{Impact of magnetic easy-axis anisotropy}
To stabilize the collinear magnetic ground state against Dzyaloshinskii-Moriya interaction (DMI), an easy-axis anisotropy
\begin{align}
    \hat{H}_{\mathrm{aniso}}
    =
    -\frac{A}{\hbar^2}
    \sum_{i}
    \qty(S_i^z)^2
\end{align}
is necessary.
Since the collinear magnetic ground state is metastable even in the presence of DMI, we did not consider it in the main text for simplicity.
Here, we demonstrate that our results hold even in the presence of an easy axis.

\begin{figure}
    \centering
    \includegraphics[width=\linewidth]{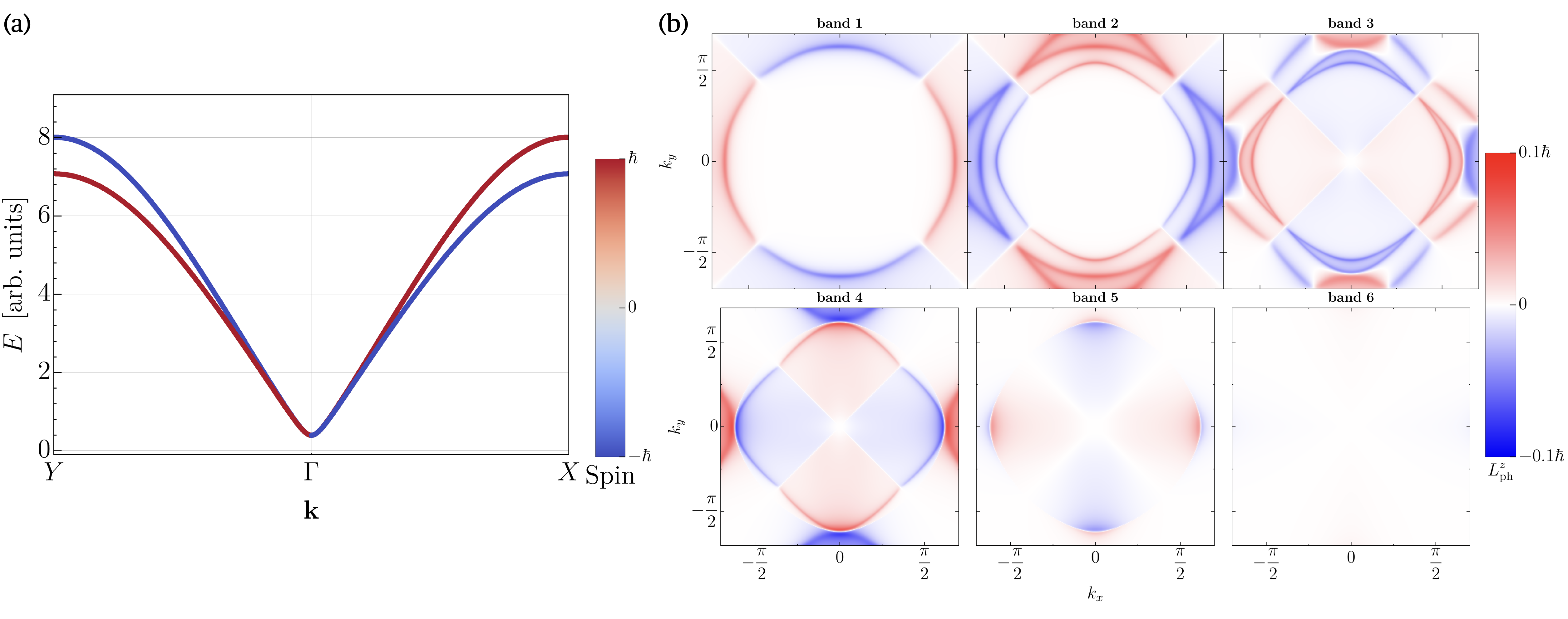}
    \caption{%
        (a) Magnon band structure of the square-lattice altermagnet with easy-axis anisotropy.
        The parameters are
        $
            J_{\mathrm{AFM}} = -1,
            J_{\mathrm{FM}_1} = 0.383,
            J_{\mathrm{FM}_2} = 0.5,
            S = 1,
            \text{ and }
            A = 0.01
        $.
        (b) Phonon angular momentum of the coupled magnon-phonon system for all bands across the entire Brillouin zone in the presence of the easy-axis anisotropy ($A = 0.01$).
        Bands are labeled in descending order starting with the highest energy.
        Additional parameters are equal to those in the main text.%
    }
    \label{fig:easy_axis}
\end{figure}
An easy-axis anisotropy enters the harmonic magnon Hamiltonian on the diagonal elements as an on-site energy shift that is independent of $\vect{k}$:
\begin{align}
    \hamil_{\mathrm{aniso}}^{(2)}
    &=
    2 A S
    \sum_{n \vect{k}}
    \adj{a}_{n \vect{k}}
    a_{n \vect{k}}
    =
    A S
    \sum_{\vect{k}}
    \adjvect{\alpha}_{\vect{k}}
    \idmatr
    \vect{\alpha}_{\vect{k}}
    +
    \mathrm{const.}
    \ ,
\end{align}
where
    $
    \adjvect{\alpha}_{\vect{k}}
    =
    (
        \adj{a}_{\uparrow,\vect{k}}\ \adj{a}_{\downarrow,\vect{k}}\ 
        a_{\uparrow,-\vect{k}}\ a_{\downarrow,-\vect{k}}
    )
$
and $\adj{a}_{n \vect{k}}$ and $a_{n \vect{k}}$ are the bosonic creation and annihilation operators of spin flips on sublattice $n = \uparrow, \downarrow$.
Under a unitary transformation $\matr{U}_{\vect{k}}$ that diagonalizes the total Hamiltonian, the addition of the easy axis would merely rigidly shift the band structure:
\begin{align}
    \adjmatr{U}_{\vect{k}}
    (A S \idmatr)
    \matr{U}_{\vect{k}}
    =
    A S \idmatr
    \ .
\end{align}
However, since a paraunitary transformation $\matr{T}_{\vect{k}}^{\mathrm{m}}$, which satisfies 
$
    \adj{(\matr{T}_{\vect{k}}^{\mathrm{m}})}
    \metricmatr
    \matr{T}_{\vect{k}}^{\mathrm{m}}
    =
    \metricmatr
$
with
$\metricmatr = \diag(\mqty{1 & 1 & -1 & -1})$, is necessary for bosonic Bogoliubov-de Gennes Hamiltonians, the band structure and the magnon wave functions are modified by the anisotropy.
This renormalization is particularly important in the vicinity of $\vect{k} = \vect{0}$, where the linear Goldstone mode is gapped out and the dispersion becomes quadratic in the long-wavelength limit [cf.~\cref{fig:easy_axis}(a)].
At the Brillouin zone boundary, the anomalous coupling terms approach zero [i.e., $\eta_{\vect{k}} \to 0$ in \cref{eq:magnon_kernel}], which means that $\matr{T}_{\vect{k}}^{\mathrm{m}}$ becomes unitary.
Hence, close to the boundary, the anisotropy mostly shifts the energy, but does not change the wave function.
Close to the center, where the anisotropy has the largest qualitative effect, the magnon-phonon coupling vanishes because it scales quadratically with $k$ in the long-wavelength limit [cf.~\cref{eq:mpc_kernel}].

We have verified the above reasoning numerically by adding an easy-axis anisotropy $A = 0.01$ to the model presented in the main text.
The phonon angular momentum texture, depicted in \cref{fig:easy_axis}(b), exhibits the same symmetry, qualitative features, and order of magnitude as in Fig.~5 in the main text.

\subsection{Impact of magnon-phonon coupling}
In the main text, the effects of one specific form of the spin-lattice coupling Hamiltonian have been analyzed. Here, we present additional results, highlighting that, while the exact kind and strength of the interactions may differ, the key features of the results remain the same. 
In the following we study the second term in \cref{eq: DMI derivative} assuming that $\pdv{D}{R} = 0$, which results in
\begin{align}
    \hat{H}_{\mathrm{DMI}}^{(1)}
    =
    \hbar S
    \sum_{i,j}
    \sum_{\alpha, \beta}
    \frac{D}{R} \left[ \kron{\alpha}{\beta} - \hat{R}^{\alpha} \hat{R}^{\beta} \right] ( S_i^{\beta} z_j - S_j^{\beta} z_i ) ( u_i^{\alpha} - u_j^{\alpha} )
    \ .
    \label{eq: coupling G=0}
\end{align}
In the following, we scrutinize two cases: (i) next-nearest neighbor interaction (i.e., $z_i = z_j$, as in the main text) and (ii) nearest neighbor interaction (i.e., $z_i = -z_j$).
In both cases, we choose the parameters
$J_{\mathrm{AFM}} = -1$,
$J_{\mathrm{FM}_1} = 0.383$,
$J_{\mathrm{FM}_2} = 0.5$,
$S = 1$,
$M = 10$,
$K_{\mathrm{L}}^{(1)} = 160$,
$K_{\mathrm{T}} = K_{\mathrm{L}}^{(2)} = 40$,
$D = 0.25$,
and
$a = 1$.

\begin{figure}
    \centering
    \includegraphics[width=0.9\linewidth]{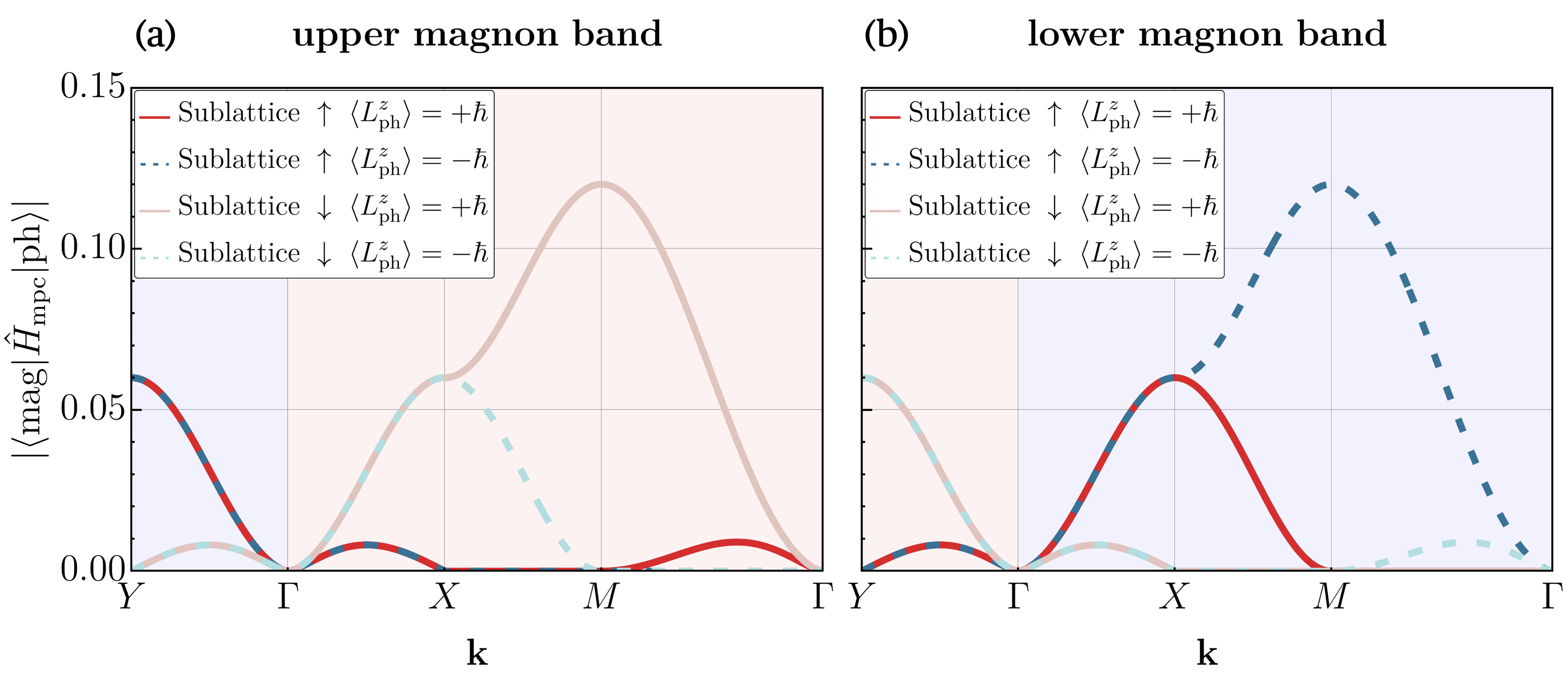}
    \caption{%
        Magnon-phonon coupling strength between the (a) upper, (b) lower magnon branch and the circularly and sublattice-polarized phonon modes.
        All lines represent the absolute value of one of the 8 matrix elements of $\hat{H}_{\mathrm{mpc}}$ [\cref{eq: coupling G=0}; only next-nearest-neighbor interactions (case (i))] represented in the basis of the magnon bands and the sublattice-resolved chiral phonons (cf.~\cref{sec:mpc_select} for mathematical details).
        The background color and the line color indicate the signs of the magnon spin and the phonon angular momentum, respectively.
        Note that the magnon bands are degenerate along $\overline{\Gamma \mathrm{M}}$, rendering their labeling arbitrary.
    }
    \label{fig:coupling_strength nnn G=0}
\end{figure}

\begin{figure}
    \centering
    \includegraphics[width=0.9\linewidth]{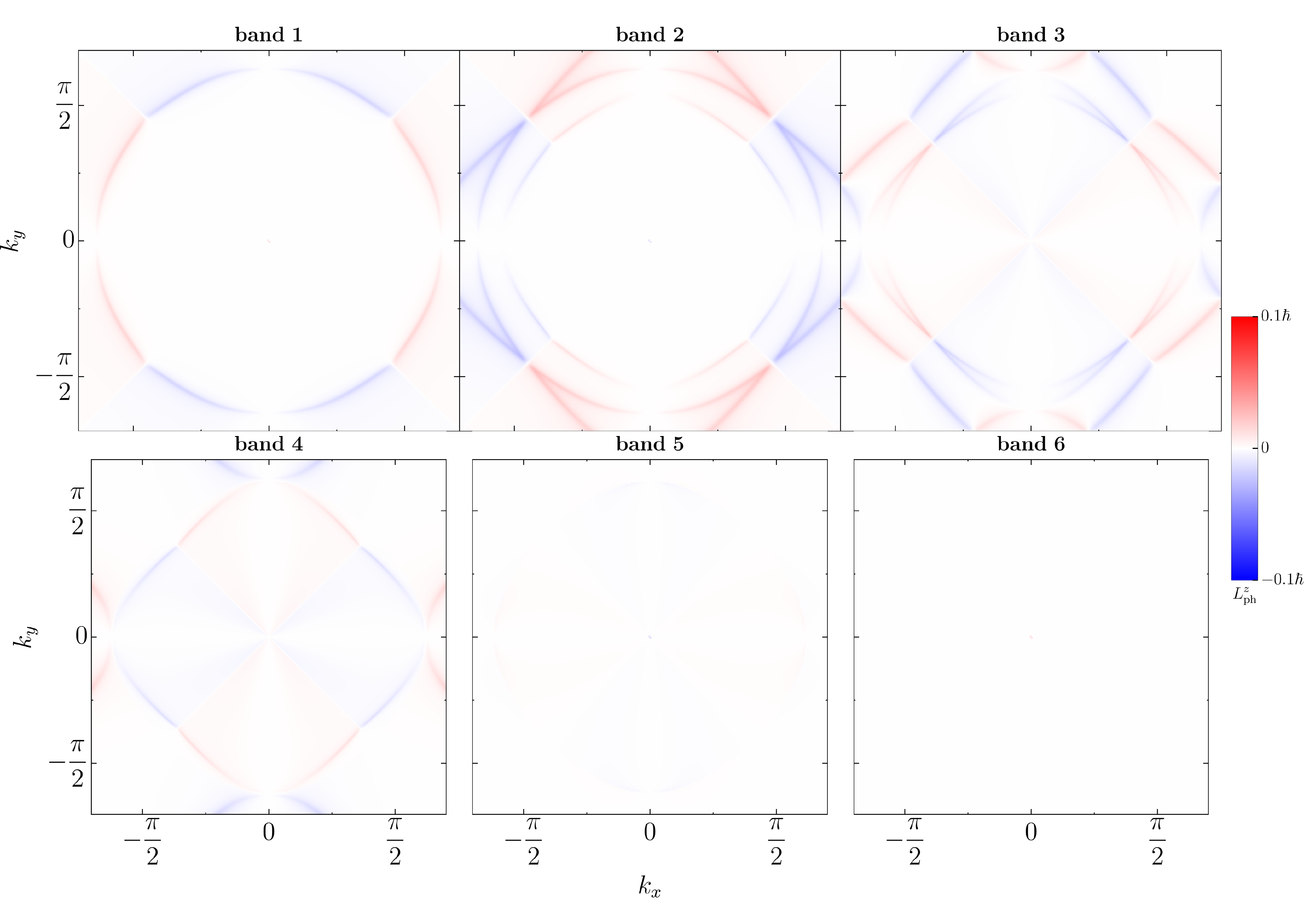}
    \caption{%
        Phonon angular momentum for all bands across the entire Brillouin zone for the spin-lattice coupling in \cref{eq: coupling G=0} considering next-nearest neighbor interactions [case (i)].
        Nonzero angular momentum occurs around regions of avoided crossings and clearly reflects the $d$-wave character of the altermagnet.
        Bands are labeled in descending order starting with the highest energy.
    }
    \label{fig:density_plot_bz nnn G=0}
\end{figure}

The results for the coupling strength and phonon angular momentum across the entire Brillouin zone for case (i) are presented in \cref{fig:coupling_strength nnn G=0} and \cref{fig:density_plot_bz nnn G=0}, respectively.
We find that along $\overline{\Gamma \mathrm{X}}$ and $\overline{\Gamma \mathrm{Y}}$ the chiral selectivity of the coupling vanishes for both magnon bands (\cref{fig:coupling_strength nnn G=0}), which results in additional \enquote{nodes} in the phonon angular momentum texture (\cref{fig:density_plot_bz nnn G=0}).
This is rooted in the fact that along those directions the magnon bands couple to linearly polarized phonons, which are equal superpositions of left- and right-handed circularly polarized lattice waves.
In contrast to the nodal lines enforced by the spin/magnetic point group symmetries of the model, the new zeros in the phonon angular momentum texture are not associated with sign changes of $\expval{L_{\text{ph}}^z}$.
The chiral selectivity is partially restored when departing from these lines such that a finite phonon angular momentum emerges.
Overall, we find that the phonon angular momentum in this case is smaller and more localized around the regions of the avoided crossings (see \cref{fig:density_plot_bz nnn G=0}).

\begin{figure}
    \centering
    \includegraphics[width=0.9\linewidth]{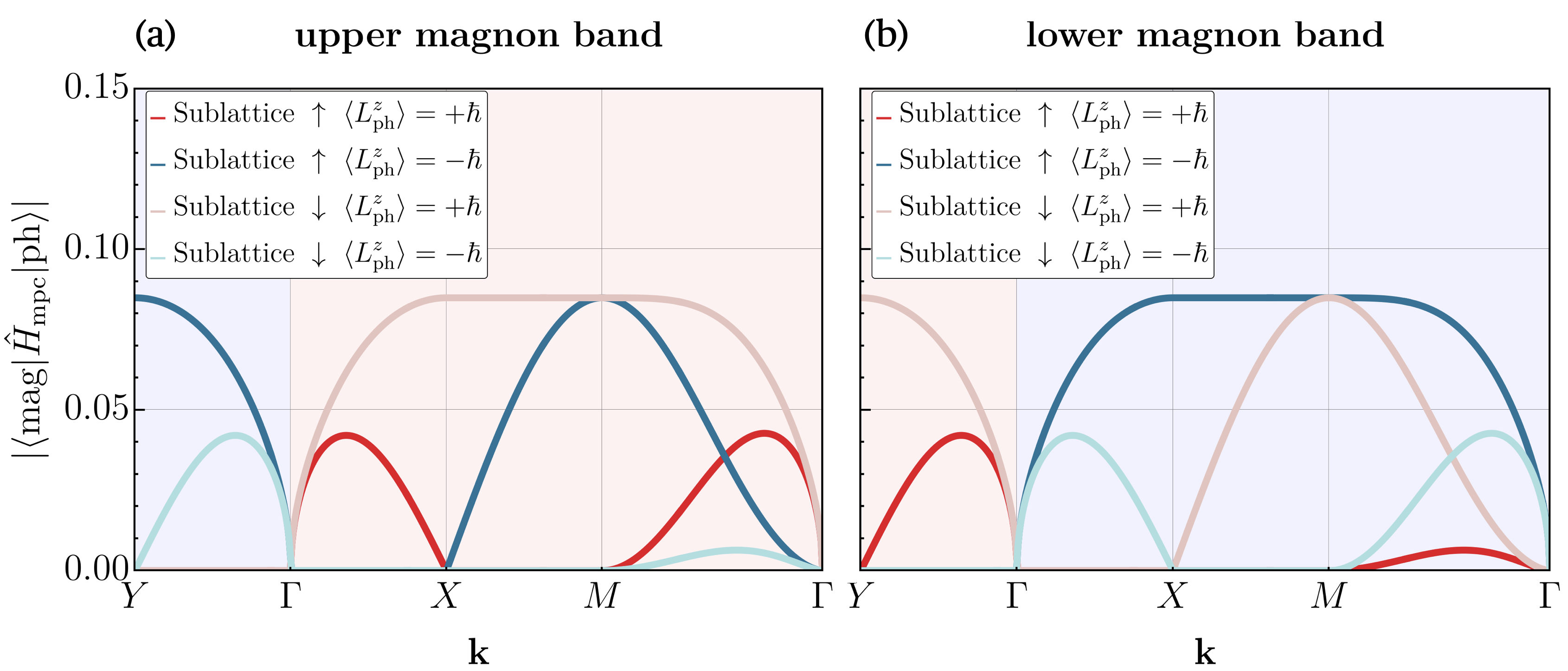}
    \caption{%
        Magnon-phonon coupling strength between the (a) upper, (b) lower magnon branch and the circularly and sublattice-polarized phonon modes.
        All lines represent the absolute value of one of the 8 matrix elements of $\hat{H}_{\mathrm{mpc}}$ [\cref{eq: coupling G=0}; only nearest-neighbor interactions (case (ii))] represented in the basis of the magnon bands and the sublattice-resolved chiral phonons (cf.~\cref{sec:mpc_select} for mathematical details).
        The background color and the line color indicate the signs of the magnon spin and the phonon angular momentum, respectively.
        Note that the magnon bands are degenerate along $\overline{\Gamma \mathrm{M}}$, rendering their labeling arbitrary.
    }
    \label{fig:coupling_strength nn G=0}
\end{figure}

\begin{figure}
    \centering
    \includegraphics[width=0.9\linewidth]{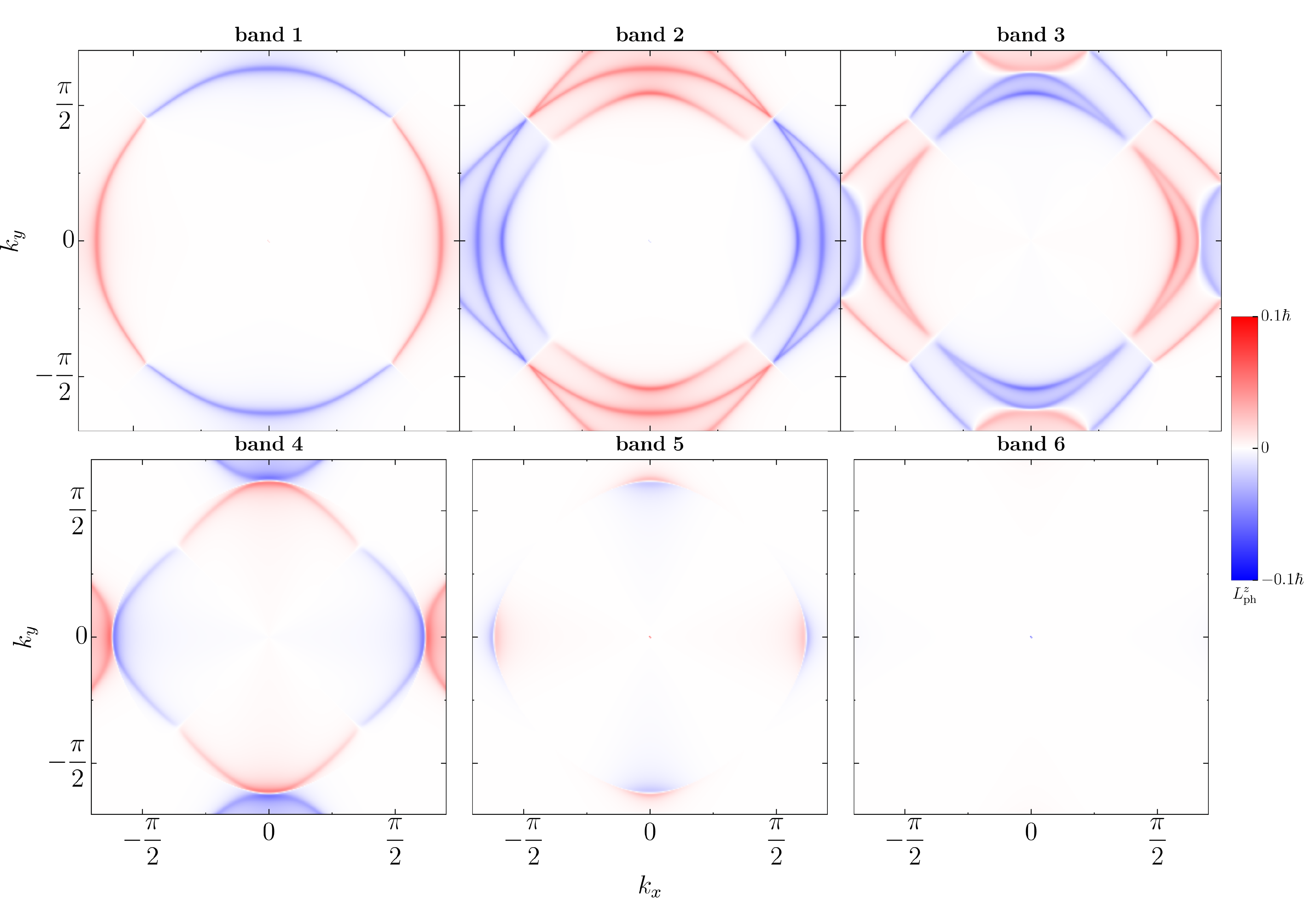}
    \caption{%
        Phonon angular momentum for all bands across the entire Brillouin zone for the spin-lattice coupling in \cref{eq: coupling G=0} considering nearest neighbor interactions [case (ii)].
        Nonzero angular momentum occurs around regions of avoided crossings and clearly reflects the $d$-wave character of the altermagnet.
        Bands are labeled in descending order starting with the highest energy.
    }
    \label{fig:density_plot_bz nn G=0}
\end{figure}

\begin{figure*}
    \centering
    \includegraphics[width=\linewidth]{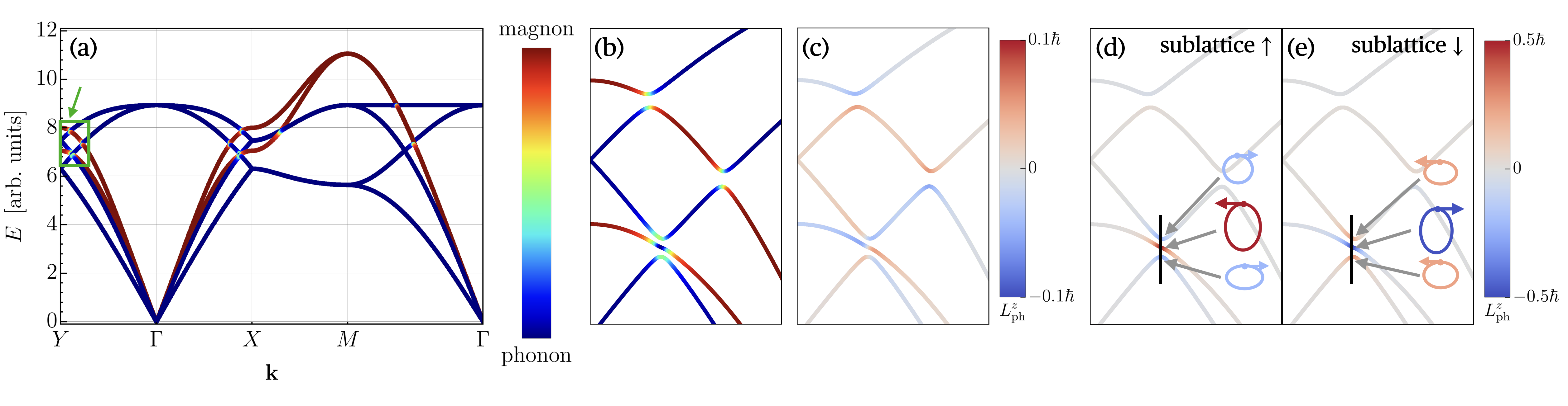}
    \caption{%
        Coupled magnon-phonon hybrid system for the spin-lattice coupling in \cref{eq: coupling G=0} considering nearest neighbor interactions [case (ii)].
        (a) Band structure of the magnon polarons along a high-symmetry path in the magnetic Brillouin zone.
        The color indicates the magnon/phonon character of the quasiparticles.
        The region highlighted by the green box and arrow is displayed in greater detail in (b)--(e).
        (b) Zoom of the avoided crossing pattern and quasiparticle character indicated by color.
        (c) Illustration of the same region but depicting the phonon angular momentum.
        (d),~(e) Sublattice-resolved phonon angular momentum for sublattice $\uparrow$ and sublattice $\downarrow$, respectively.
        The insets demonstrate the real-space vibrational motion (amplitude and phase) of the respective sublattices.
    }
    \label{fig:hybridized_system nn G=0}
\end{figure*}

\cref{fig:coupling_strength nn G=0} and \cref{fig:density_plot_bz nn G=0} display the results for case (ii) of nearest neighbor spin-lattice coupling.
Here, the coupling takes place between spins of different ground state orientations.
We find that, in comparison to case (i), the chirality-selective coupling is restored along $\overline{\Gamma \mathrm{X}}$ and $\overline{\Gamma \mathrm{Y}}$ (compare \cref{fig:coupling_strength nn G=0} to \cref{fig:coupling_strength nnn G=0}).
Overall, the features of the coupling strength depicted in \cref{fig:coupling_strength nn G=0} are reminiscent of those in Fig.~3 of the main text.
The chiral selectivity along $\overline{\mathrm{X M}}$ is incomplete due to nonzero coupling to the opposite chirality on the other sublattice (with zero spin precession amplitude).
This feature fundamentally sets the nearest neighbor spin-lattice coupling apart from the previous calculations, which involved next-nearest neighbor spin-lattice coupling.
The latter only installed an \emph{intra-sublattice} coupling between magnons and phonons.
This is different for the spin-lattice coupling considered here, which is an \emph{inter-sublattice} coupling.
The magnon does not have to be localized on the sublattice of the phonons because the dynamics of spins on one sublattice is directly coupled to the dynamics of nearest-neighbor atoms (of the other sublattice) via the inter-site part $\vect{u}_i \cdot \vect{S}_j$.
There are still intra-sublattice coupling terms because of the inevitable on-site part $\vect{u}_i \cdot \vect{S}_i$ of the magnon-phonon coupling.
Note that the additional features along $\overline{\Gamma \mathrm{M}}$ are irrelevant as the distiction of the two magnon bands along the nodal lines of the altermagnet is arbitrary.
Another difference to Fig.~3 is the increased coupling to the other sublattice (i.e., the one exhibiting lower spin precession amplitudes) along $\overline{\Gamma \mathrm{X}}$ and $\overline{\Gamma \mathrm{Y}}$.
This yields an increased ellipticity of the hybridized eigenmodes compared to Fig.~4, which we show in \cref{fig:hybridized_system nn G=0}.

In summary, both \cref{fig:density_plot_bz nnn G=0} and \cref{fig:density_plot_bz nn G=0} exhibit the $d$-wave phonon angular momentum character, thereby further supporting the results of the main manuscript.

\FloatBarrier
\bibliography{refs}